\documentclass[12pt,preprint]{aastex}

\newcommand{\bc}{\begin{center}}
\newcommand{\ec}{\end{center}}
\newcommand{\be}{\begin{equation}}
\newcommand{\ee}{\end{equation}}
\newcommand{\ba}{\begin{eqnarray}}
\newcommand{\ea}{\end{eqnarray}}
\newcommand{\bt}{\begin{tabular}}
\newcommand{\et}{\end{tabular}}

\newcommand{\rxte}{{\sl RXTE}}

\def\farcs{\hbox{$.\!\!^{\prime\prime}$}}
\def\farcm{\hbox{$.\!\!^{\prime}$}}

\def\cxo{{\sl CXO}}
\newcommand{\sgr}{\mbox{Swift\,J1834.9$-$0846 }}
\newcommand{\sgrnos}{\mbox{Swift\,J1834.9$-$0846}}
\def\xmm{{\sl XMM-Newton}}

\def\nh{N_{\rm H}}

\def\cgs{erg s$^{-1}$ cm$^{-2}$}

\def\farcs{\hbox{$.\!\!^{\prime\prime}$}}
\def\farcm{\hbox{$.\!\!^{\prime}$}}

\begin{document}

\title{X-ray observations of a new unusual magnetar \sgrnos}

\author{
Oleg\  Kargaltsev\altaffilmark{1}\altaffiltext{1}{Dept.\ of Astronomy, University of Florida, Bryant Space Science Center, Gainesville, FL 32611; oyk100@astro.ufl.edu},  
Chryssa\ Kouveliotou\altaffilmark{2}\altaffiltext{2}{Science \& Technology Office, ZP12, NASA/Marshall Space Flight Center, Huntsville, AL 35812, USA},
George\ G.\ Pavlov\altaffilmark{3,4}\altaffiltext{3}{Pennsylvania State University, 525 Davey Lab, University Park, PA 16802, USA}\altaffiltext{4}{St.-Petersburg State Polytechnical University, Polytekhnicheskaya ul.\ 29, 195251, St.-Petersburg, Russia},
Ersin\ G\"o\u{g}\"u\c{s}\altaffilmark{5}\altaffiltext{5}{Sabanc\i University, Faculty of Engineering and Natural Sciences, Orhanl\i$-$Tuzla, \.{I}stanbul 34956, Turkey},
Lin\ Lin\altaffilmark{5},
Stefanie\ Wachter\altaffilmark{6}\altaffiltext{6}{Infrared Processing and Analysis Center, California Institute of Technology, Pasadena, CA 91125},
Roger\ L.\ Griffith\altaffilmark{6},
Yuki\ Kaneko\altaffilmark{5},
 George\ Younes\altaffilmark{2,7}\altaffiltext{7}{Universities Space Research Association, 6767 Old Madison Pike, Suite 450, Huntsville, AL 35806, USA}}

\begin{abstract}

We present X-ray observations of the new transient magnetar 
Swift\,J1834.9--0846,  
discovered with {\sl Swift} BAT on 2011 August 7. The data were obtained with 
{\sl Swift}, {\sl RXTE}, {\sl CXO}, and {\sl XMM-Newton} both before and after the outburst. Timing analysis reveals single peak pulsations with a period of 2.4823 s  and  an unusually high pulsed fraction, $85\%\pm10\%$.  Using the {\sl RXTE} and {\sl CXO} data, we estimated the period derivative, $\dot{P}=8\times 10^{-12}$ s s$^{-1}$,
 and confirmed the high magnetic field of the source, $B=1.4\times 10^{14}$ G.
The decay of the persistent X-ray flux, spanning 48 days, is consistent with a power law, $F\propto t^{-0.5}$. In the {\sl CXO}/ACIS image, we find that the highly absorbed point source is surrounded by extended emission, which most likely 
is a dust scattering  halo. 
 Swift J1834.9--0846 is located near the center of the radio supernova remnant W41 and TeV source HESS J1834--087.  An association with W41 would imply a source distance of about 4 kpc; however, any relation to the HESS source remains unclear, given the presence of several other 
 candidate counterparts for the latter source in the field. Our search for an IR counterpart of Swift J1834.9--0846 revealed no source down to $K_{s} \sim 19.5$ within the 0\farcs6 {\sl CXO} error circle. 
 
\end{abstract}


\section{Introduction}

The population of magnetars has been  growing rapidly in the last five years, reaching 24 objects as of August 2011.
  Originally comprised of Soft Gamma Repeaters (SGRs) and Anomalous X-ray Pulsars (AXPs) \citep{woods2006}, the magnetar population now includes a few more neutron star (NS) groups that have been acknowledged as magnetar candidates. Most of these NSs are slow rotators emitting multiple, very short (a few times
 100\,ms) hard X-ray/soft $\gamma$-ray bursts. Their X-ray luminosities are likely powered by the decay of their high magnetic fields (up to $B\sim10^{15}$ G), rather than rotational energy losses due to their gradual spin-down \citep{paczynski1992,duncanthompson1992,thompsonduncan1995,thompsonduncan1996}. The current synergy between NASA's three observatories ({\sl RXTE}, {\sl Swift}, and {\sl Fermi}) has enabled a much higher rate of discovery of these objects in the last three years. During July - August 2011 alone, two new candidate magnetars were discovered in X-rays, Swift\,J1822.3$-$1606 and \sgrnos, when they triggered the {\sl Swift}/Burst Alert Telescope (BAT) and the {\sl Fermi}/Gamma-ray Burst Monitor (GBM). Their timing properties were subsequently established with {\it Rossi X-ray Timing Explorer} ({\sl RXTE}) observations, clinching their magnetar nature. We report here on the X-ray spectral and temporal properties of the latter source. 

\sgr was discovered on 2011 August 7, when a soft, short burst from the source triggered the BAT at 19:57:46 UT \citep{delia2011}; approximately 3.3 hours later, at 23:16:24.91 UT, another SGR-like burst triggered GBM from the general direction of the earlier BAT location \citep{guiriec2011}. Although the GBM location included a large area with several magnetar sources, the near time coincidence and the X-ray properties of these events pointed to a common origin of a new source \citep{barthelmy2011}. The source triggered the BAT again on 2011 August 30 at 23:41:12 UT \citep{hoversten2011}.

Optical observations of the field $\sim16$\,min after the BAT trigger with the Special Astrophysical Observatory (SAO)/Big Telescope Alt-azimuth (BTA) 6-m telescope detected an object at magnitude $R_{\rm c}=23.44\pm0.34$ \citep{moskvitin2011}. Simultaneous observations with the 1.5-m Observatorio de Sierra Nevada (OSN) telescope in the I band did not detect that object to a limit of $I=21.6$ \citep{tello2011}. Archival IR images of the region as part of the UKIDSS Galactic Plane Survey \citep{lucas2008}  in the J, H and K bands on 2007 May 10, revealed two sources close to the {\sl Swift}/X-ray Telescope (XRT) location of \sgr \citep{levan2011}. None of these objects coincided with the very precise X-ray position subsequently derived from our {\sl Chandra} Target of Opportunity (ToO) observation \citep{gogus2011b}. 

{\sl RXTE}/PCA observations of the source on 2011 August 9-10 detected a coherent pulsation at $\nu =0.402853(2)$ Hz, which corresponded to a spin period 
$P=2.482295$ s \citep{gogus2011a}; this result was later confirmed with our {\sl Chandra} ToO observation on 2011 August 22 \citep{gogus2011b}. Continuous {\sl RXTE} monitoring of the source over a time span of two weeks revealed a spin-down rate $\dot{\nu}= -1.3(2)\times10^{-12}$ Hz s$^{-1}$ \citep{kuiper2011}. The corresponding estimate of the surface 
 magnetic field, $B=1.4\times10^{14}$ G, confirmed the magnetar nature of \sgrnos. 
 
\sgr is  located in  a field  rich with high-energy sources, which include SNR W41 \citep{shaver1970,tian2007}, the TeV source HESS J$1834$-$087$ \citep{aharonian2005},  the GeV source 2FGL~J1834.3$-$0848 \citep{abdo2011}, and the PSR/PWN candidate  XMMU J183435.3$-$84443/CXOU J183434.9$-$084443  \citep{mukherjee2009,misanovic2011}. Attempts to understand the nature and relations between these  sources had already prompted X-ray observations  with \cxo\/ and \xmm\/ before the discovery of \sgr \citep{mukherjee2009,misanovic2011}. We have triggered additional observations of the region with both \cxo\/ and \xmm. Here we describe  the analyses of the {\it RXTE}, {\it Swift}, {\it Fermi}, and \cxo\/ data and compare them to the earlier observations. The  \xmm\/  results will be reported in a separate paper. Section \ref{sec:xray} describes the data sets presented here, and Section \ref{sec:location} presents the \cxo\/ location and discusses possible optical counterparts.  We present the lightcurve of the persistent emission in Section  \ref{sec:variab} and the results of our timing and spectral analyses in Sections \ref{sec:timing} and \ref{sec:spectra}, respectively.   
  Finally, we compare the
properties of \sgr with those of other magnetars and  discuss the possible relation of \sgr to other sources in the field  in Section \ref{sec:discussion}. 

\section{X-ray observations and data reduction.}
\label{sec:xray}

The field of \sgr was observed in X-rays on 29  occasions with several telescopes; the majority was in 2011, with two earlier observations in 2005 and 2009 (see Table \ref{tab:xraylist}). We have analyzed here 20 {\it Swift}/XRT observations, 8 {\it RXTE}/PCA observations, and one \cxo/ACIS observation.

\subsection{{\it Swift}/XRT data} 

Of the 20 {\sl Swift}/XRT observations  
listed in Table \ref{tab:xraylist}, four were carried out in the Photon Counting (PC) mode and sixteen in the Window Timing (WT) mode which provides much better temporal resolution (1.8 ms) at the expense of imaging. We used the HEASOFT\footnotemark \footnotetext{Version 6.10, \url{http://heasarc.gsfc.nasa.gov/docs/software/lheasoft/}} analysis tools to reduce and analyze the data. We extracted spectra from the Level 2 event data using the standard grade selection of 0--12 and 0--2 for the PC and WT mode data, respectively. For the PC mode data, we used  an $r=15''$ circle as the source region and an annulus with the same center and inner and outer radii of $30''$ and $45''$ as the background region.  For the WT mode data, we extracted the source spectra using a box centered on the \cxo\/ location with a length of $30''$ aligned to the 1D image. The background spectra were extracted with a similar size box centered far away from the source. We then generated the ancillary response files with {\it xrtmkarf} for each spectrum and regrouped the source spectra with a minimum of 15 counts per bin. The spectral fitting was done in XSPEC 12.6.0. Since the source was relatively bright at the onset of the outburst episode, the first XRT observation in PC mode (performed during two separate spacecraft orbits) was split into two parts to uncover early spectral variations.  Three observations in WT mode  and three observations in PC mode were too short 
  to allow   determination of spectral parameters. These observations were, therefore, excluded from our spectral analysis.  
 
\subsection{{\it RXTE}/PCA data}

\sgr was observed with \rxte\/ in eight pointings with a total exposure time of about 50\,ks spanning over 30\,days (see Table \ref{tab:xraylist}). The \rxte\/ data were collected with the Proportional Counter Array (PCA; \cite{jahoda1996}) operating with two out of the five available proportional counter units in most of the observations. All data were collected in the GoodXenon mode, where each photon is time tagged with a minimum time resolution of about 1\,$\mu$s. We used the PCA data primarily  for timing analysis as it is not an imaging instrument, and the source intensity is relatively dim compared to the bright background X-ray emission (e.g., diffuse Galactic ridge emission and bright point sources in the $1^{\circ}$ field of view of {\it RXTE}). However, we extracted the pulse peak spectrum using the longest \rxte\/ pointing to investigate the source spectral behavior in a joint PCA and \cxo\/ analysis (see Section \ref{sec:cxo-rxte}).

\subsection{\cxo\/ data}

We observed \sgr on 2011 August 22 with the \cxo\/ Advanced CCD Imaging Spectrometer (ACIS) operated in the Timed Exposure mode. The target was imaged near the aim point on the S3 chip  using 1/8 subarray 
($8'\times1'$ field of view). The data of an archival \cxo\/ observation \citep[see][for a description]{misanovic2011} were also analyzed, taking into account the different angular resolution and sensitivity. In our analysis we worked with the pipeline-produced Level 2 event files (with standard filtering applied) and utilized CIAO 4.3 with CALDB 4.4.5. The spectral fitting was done in XSPEC 12.6.0. 

\section{Source Location and Optical Counterpart Search}
\label{sec:location}

\begin{figure}
 \centering
\includegraphics[width=5.3in,angle=0]{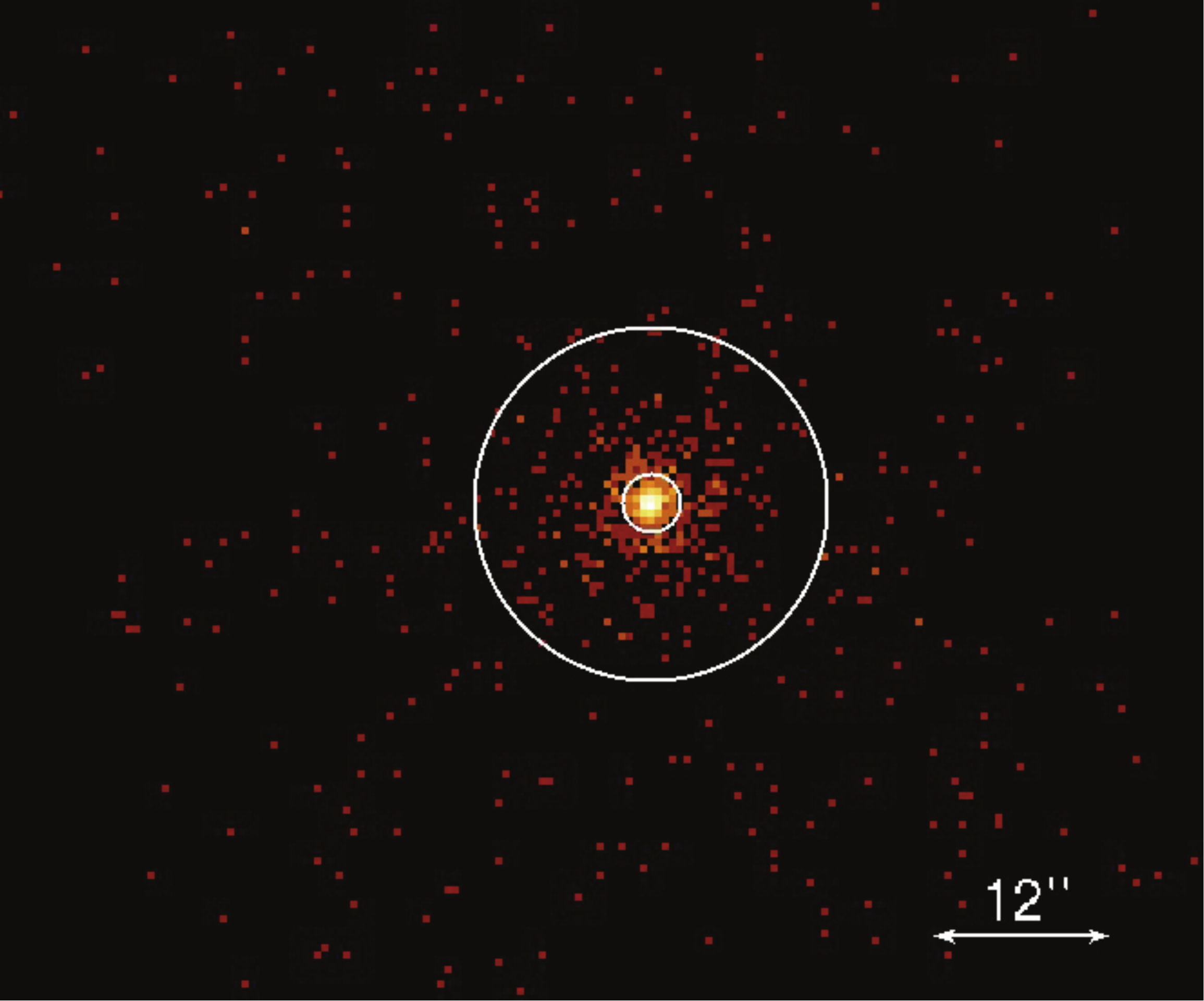}
\caption{Image of  CXOU~183452.1--084556 and surrounding emission ($0.7-10$ keV) obtained with ACIS-S3 on 2011 August 22. The radii of the inner and outer circles are $2''$ and $12''$, respectively.}
\label{fig:img:acis:ddt}
\end{figure}

We used the {\em wavdetect} CIAO tool to determine the point sources in our \cxo\/ observation. In the vicinity of the {\sl Swift}/XRT location we find a point source, which we designate CXOU~J183452.1$-$084556, centered at ${\rm R.A.}=18^{\rm h}$  34$^{\rm m}$ 52$^{\rm s}$.118, ${\rm decl.}=-08^o$ 45$'$ 56\farcs02. We also notice the presence of extended emission, up to $\simeq15''$ from the point source, with isotropic surface brightness distribution (see Section 6.2.2). The uncertainty of this position is dominated by the \cxo\/ absolute position uncertainty of $0\farcs 6$ (at 90\% confidence level).\footnote{See \url{http://cxc.harvard.edu/cal/ASPECT/celmon/}.} The \cxo\/ image of the vicinity of \sgr  is shown in Figure~\ref{fig:img:acis:ddt}.  

\begin{figure}
 \centering
 \includegraphics[width=5.3in,angle=0]{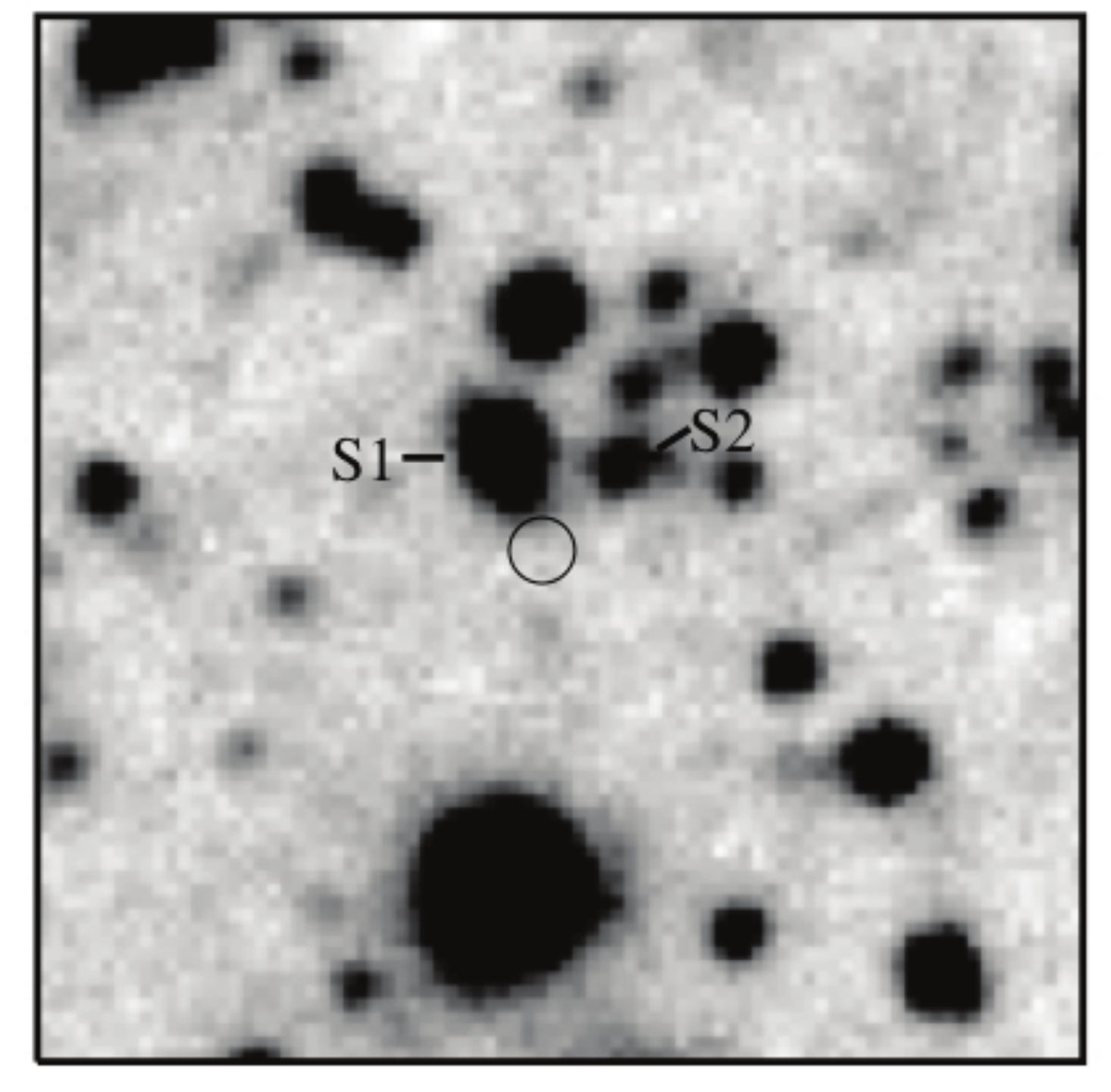}
\caption{Palomar/WIRC $K_{\rm s}$-band image showing the $r=0.6''$ \cxo\/ error circle for CXOU~183452.1--084556. The sources designated as S1 and S2  are the ones reported by \citet{levan2011}.}
\label{fig:wirc}
\end{figure}

We compared the \cxo\/ image to the archival 2MASS images of the same region of sky. We do not detect any near-infrared (NIR) sources within $2''$ distance from the position of CXOU~J$183452.1-084556$. We also observed the field of  \sgr with the Wide Field Infrared Camera \citep[WIRC; ][]{wilson2003} on the 5-m Palomar Hale telescope on 2011 August 23. WIRC has a field of view of $8\farcm 7\times 8\farcm 7$ and a pixel scale of 0.2487  arcseconds per pixel. We obtained seven dithered $K_{\rm s}$ band images, consisting of four co-added 30-second exposures taken at each dither position. 
  The atmospheric conditions were very good, with seeing $\lesssim1''$ and clear skies. The individual frames were reduced in the standard manner using IRAF, calibrated, and mosaiced together. 
The resulting image was astrometrically calibrated using 2MASS. The astrometric solution carries a formal 1$\sigma$ error of $0\farcs 1$ for the transfer of the 2MASS reference frame to the WIRC image shown in Figure \ref {fig:wirc}.   No sources are detected within the \cxo\/ error circle down to a limiting magnitude of $K_{\rm s}\sim19.5$ (at the 5$\sigma$ level). The sources designated as S1 and S2 on the figure are the ones reported earlier by \citet{levan2011}.

\section{Persistent X-ray lightcurve of \sgrnos}
\label{sec:variab}

\begin{figure}
 \centering
\includegraphics[width=7.0in,angle=0]{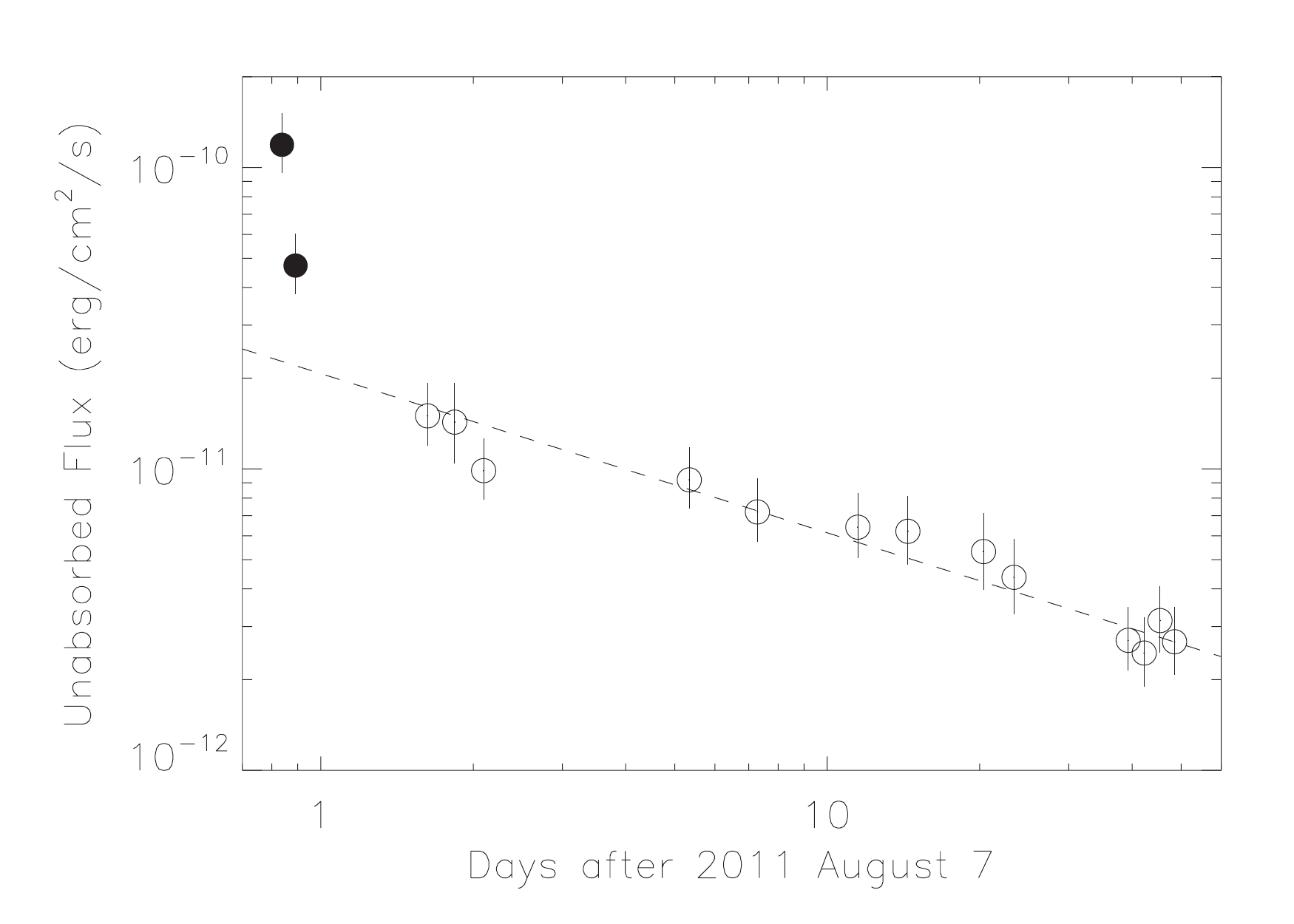}
\caption{Persistent X-ray lightcurve ($2-10$\,keV) of \sgrnos/CXOU~J183452.1--084556
 obtained from 48 days monitoring of the source with {\sl Swift}/XRT. 
The dashed line shows the best-fit power-law temporal decay model ($\propto t^{-0.53} $).   }
\label{fig:lcurve}
\end{figure}

\sgr was observed on 20 occasions with {\it Swift} after the outburst onset (see Table 1). This coverage allows us to construct a lightcurve of the source, which spans 48 days. In Figure \ref{fig:lcurve}, we present the persistent X-ray flux history in the 2$-$10\,keV range as calculated using the power-law (PL) spectral model described in Section \ref{sec:xrtspec}. The X-ray lightcurve of the source indicates a rapid decay in the very early episode ($\lesssim$ 1 day), and it is consistent with a steady flux decay over the longer term. A PL fit to the temporal decay trend (i.e., $F\propto t^{-\alpha}$) yields a good fit with $\alpha=0.53\pm 0.03$ and $\alpha=0.53\pm 0.07$ for the observed and unabsorbed fluxes,
respectively. Notice that because of the limited spatial resolution, the XRT data 
 include both the point source and the surrounding extended emission. As a consequence, the decay trend of the point source cannot be unambiguously determined from these data.

\section{Timing Analysis}
\label{sec:timing}

\subsection{\rxte}

\sgr was observed by \rxte\/ on eight occasions with a total exposure time of $\sim50$\,ks, spanning a time baseline of over 30 days (see Table \ref{tab:xraylist}). For our timing analysis we used data collected in the 2$-$10\,keV range. For each observation, we first inspected the lightcurve with 0.03125\,s time resolution and filtered out the times of short spikes and instrumental artifacts. We then converted the event arrival times to that of the Solar System Barycenter in Barycentric Dynamical Time (TDB) using the JPL DE200 ephemeris and the {\it Swift}-derived coordinates of the source. 

\begin{figure}
\centerline{
\plotone{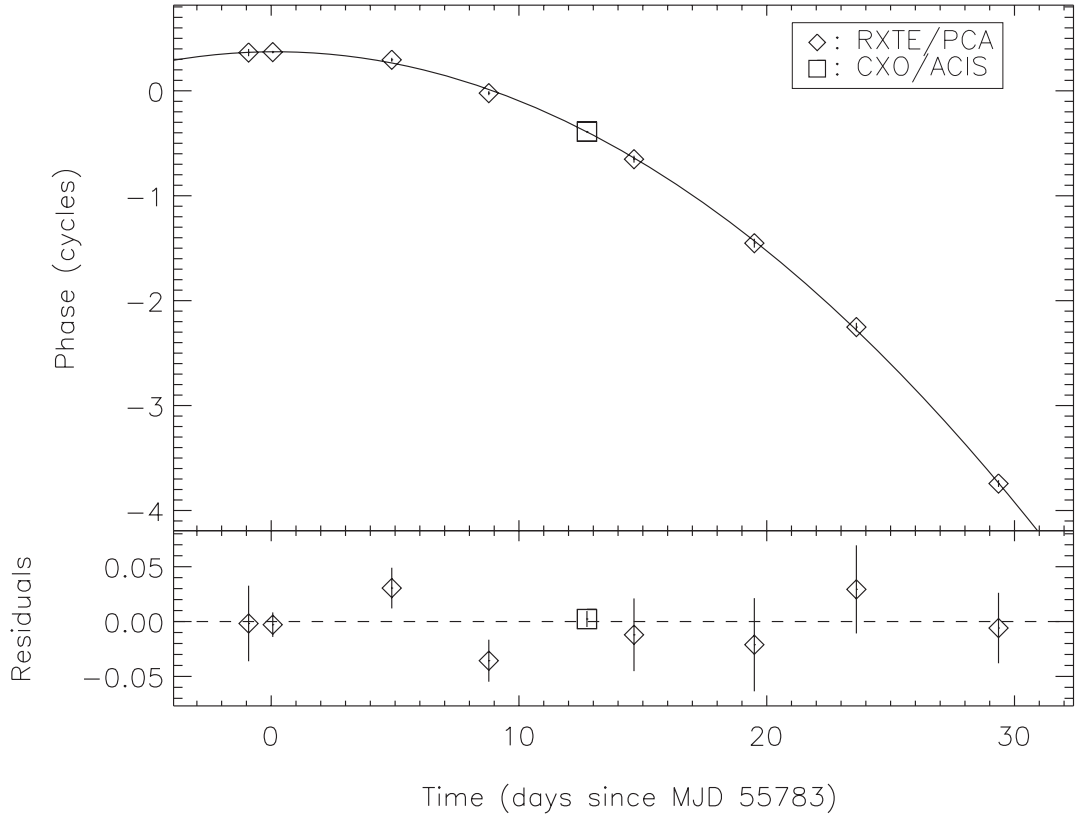}}
\caption{{\it Top panel:} Plot of phase shifts for each \rxte\/ observation of \sgrnos. The solid line is a quadratic trend that fits the time evolution of the phase shifts. {\it Bottom panel:} Residuals of the fit.}
\label{fig:spin_eph}
\end{figure}

Next, we employed a Fourier based pulse profile folding technique to determine the spin ephemeris of \sgrnos. We first generated a template pulse profile by folding the longest PCA observation (Observation ID: 96434-01-02-00) at the pulse frequency determined with a $Z_1^2$ search \citep{buccheri1983}.
 Then, we generated pulse profiles for all PCA observations as well as for the \cxo\/ pointing, and cross-correlated them with the template profile to determine the phase shifts with respect to the template. We obtain the spin ephemeris of the source by fitting the phase shifts with a first or higher order polynomial. We find that the phase drifts of \sgr are best described with a second order polynomial ($\chi^2$=7.3 for 7 degrees of freedom [dof]) that yields a spin period $P=2.4823018(1)$ s  and a period derivative $\dot{P}=7.96(12) \times 10^{-12}$ s s$^{-1}$ (epoch: 55783 MJD). In Figure \ref{fig:spin_eph} we present the drift of the pulse phase with respect to the template and the quadratic trend curve (upper panel), and the fit residuals in cycles (lower panel). The measured values of $P$ and $\dot{P}$ correspond to the following spin-down parameters: 
age $\tau= P/2\dot{P} = 4.9$ kyr, 
power $\dot{E}=4\pi^2I\dot{P}P^{-3}=2.1\times 10^{34}$ erg s$^{-1}$,
and magnetic field $B=3.2\times 10^{19} (P\dot{P})^{1/2} = 1.4\times 10^{14}$ G. 

\begin{figure}
 \centering
 \includegraphics[width=6.3in,angle=0]{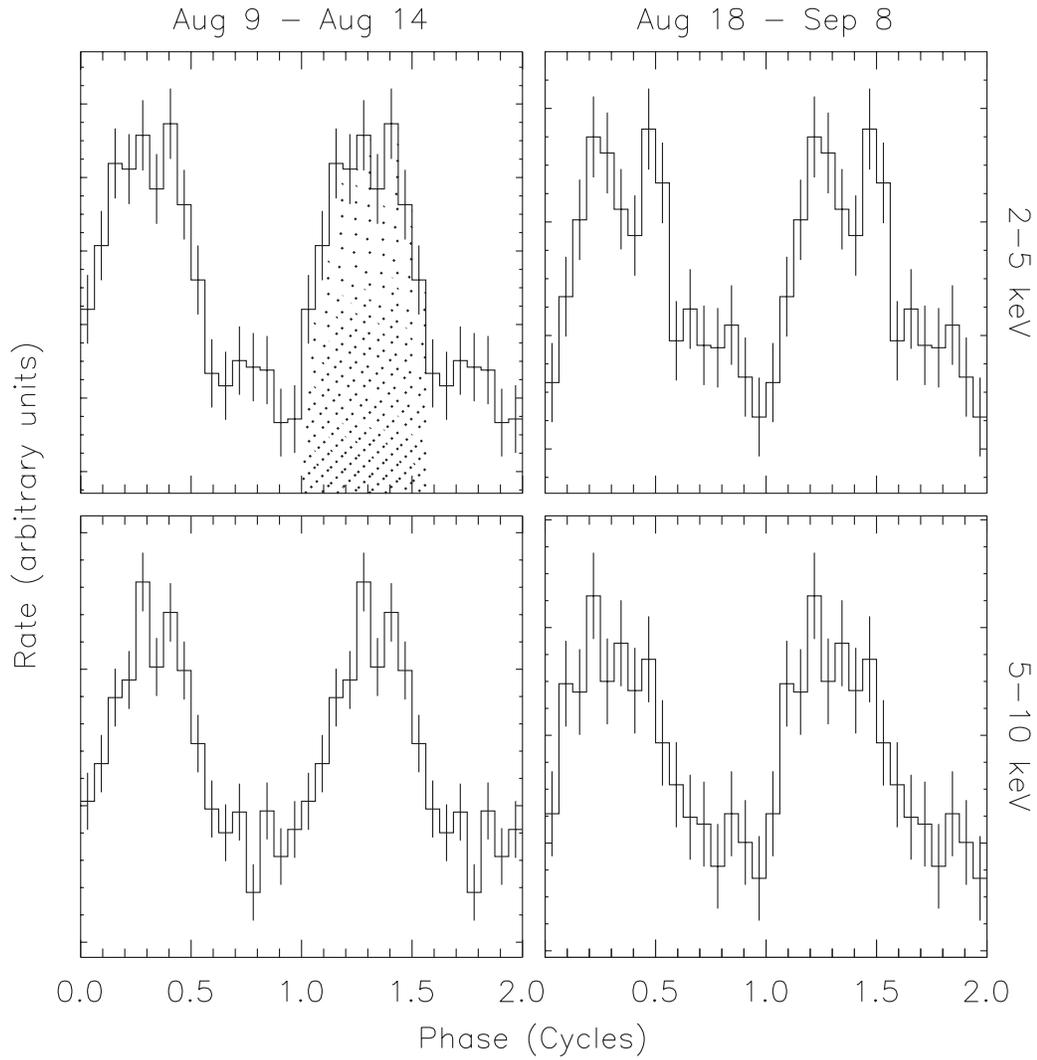}
\caption{Dependence of the \rxte\/ pulse profiles on time and energy. The shaded area in the top left panel corresponds to the phase interval used for spectral analysis (see Section \ref{sec:cxo-rxte}).}
 \label{fig:pulse:rxte}
\end{figure}

Finally, Figure \ref{fig:pulse:rxte} shows the pulse profiles obtained from several \rxte\/ observations folded together using the derived ephemeris. We note the appearance of additional harmonics in the low energy pulse profile of the source ($2$--$5$\,keV) in \rxte\/ data taken at later times.

\subsection{\cxo}

We searched for pulsations in the \cxo/ACIS data obtained in the 2011 August 22 observation. We used the 733 counts extracted from the $r=1''$ circle around the CXOU~J183452.1$-$084556 position, in the $2$--$10$\,keV
band (there are only 4 counts below 2\,keV, likely from the 
background). The time resolution in this observation  
 was 0.44104\,s (0.4\,s frame time plus 0.04104\,s charge transfer time). The photon arrival times were transformed to the Solar System Barycenter using the CIAO {\it axbary} tool. The ACIS observation started at epoch 55795.6489 MJD and continued for $T_{\rm span}=13.02$\,ks. 

We calculated the $Z^2_{1}$ statistic as a function of trial frequency
 with a step of $0.35~\mu$Hz (which is about $0.05T_{\rm span}^{-1}$) and found the maximum $Z^2_1 = 467$ at $\nu = 0.4028512~\rm{Hz} \pm 2.0~\mu \rm{Hz}$\footnote{The $1\sigma$ uncertainty is calculated as $\delta \nu= 3^{1/2}\pi^{-1}T_{\rm span}^{-1}(Z_{1,{\rm max}}^{2})^{-1/2}$ \citep[see][]{chang2011}.},
 implying a very high significance of the pulsed signal.
 We also calculated $Z^2_{n}$ for $n>1$ but did not find a strong contribution
of higher harmonics.

\begin{figure}
 \centering
\includegraphics[width=4.1in,angle=90]{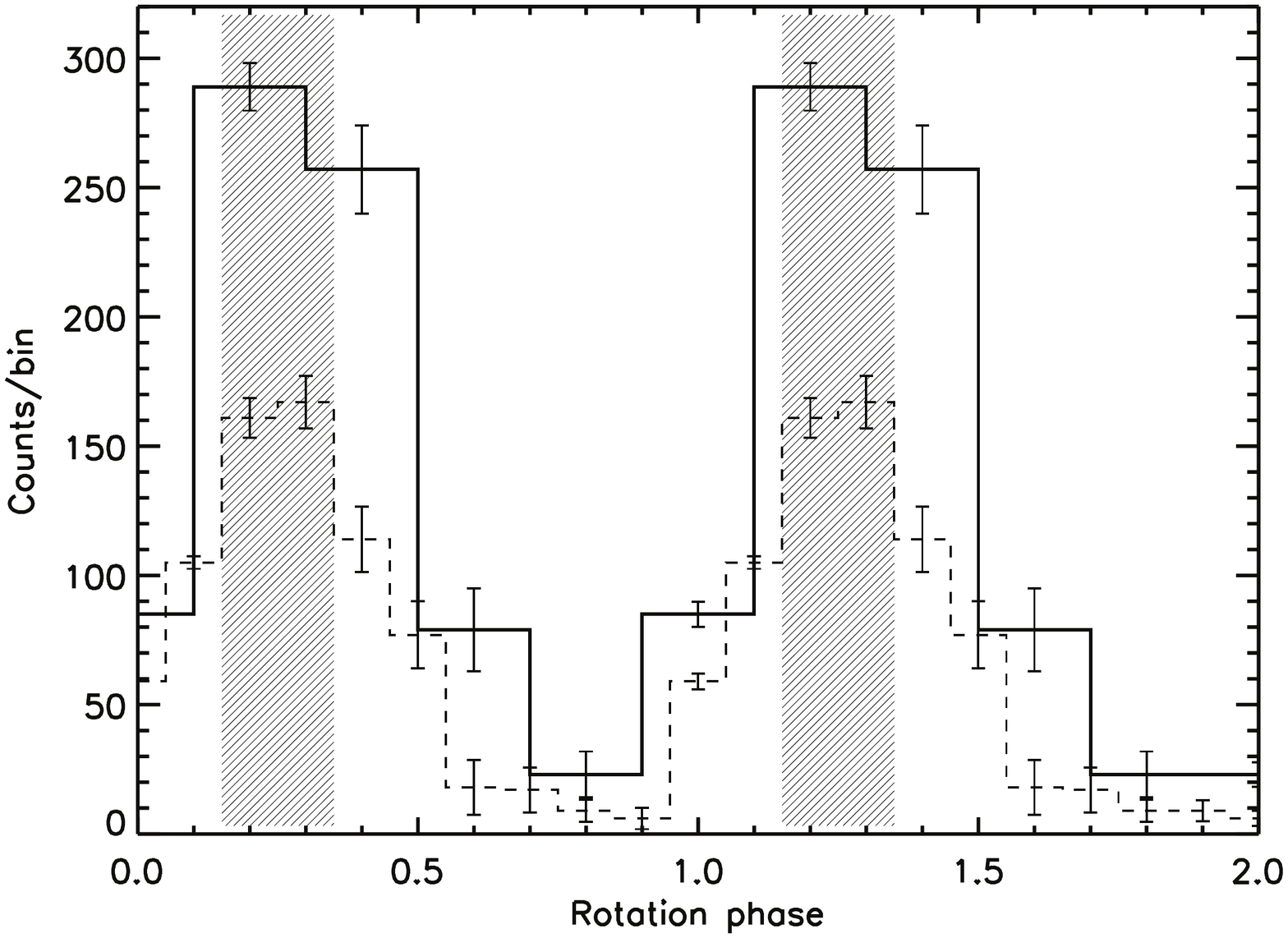}\vspace{-2.0cm}
 \includegraphics[width=4.1in,angle=90]{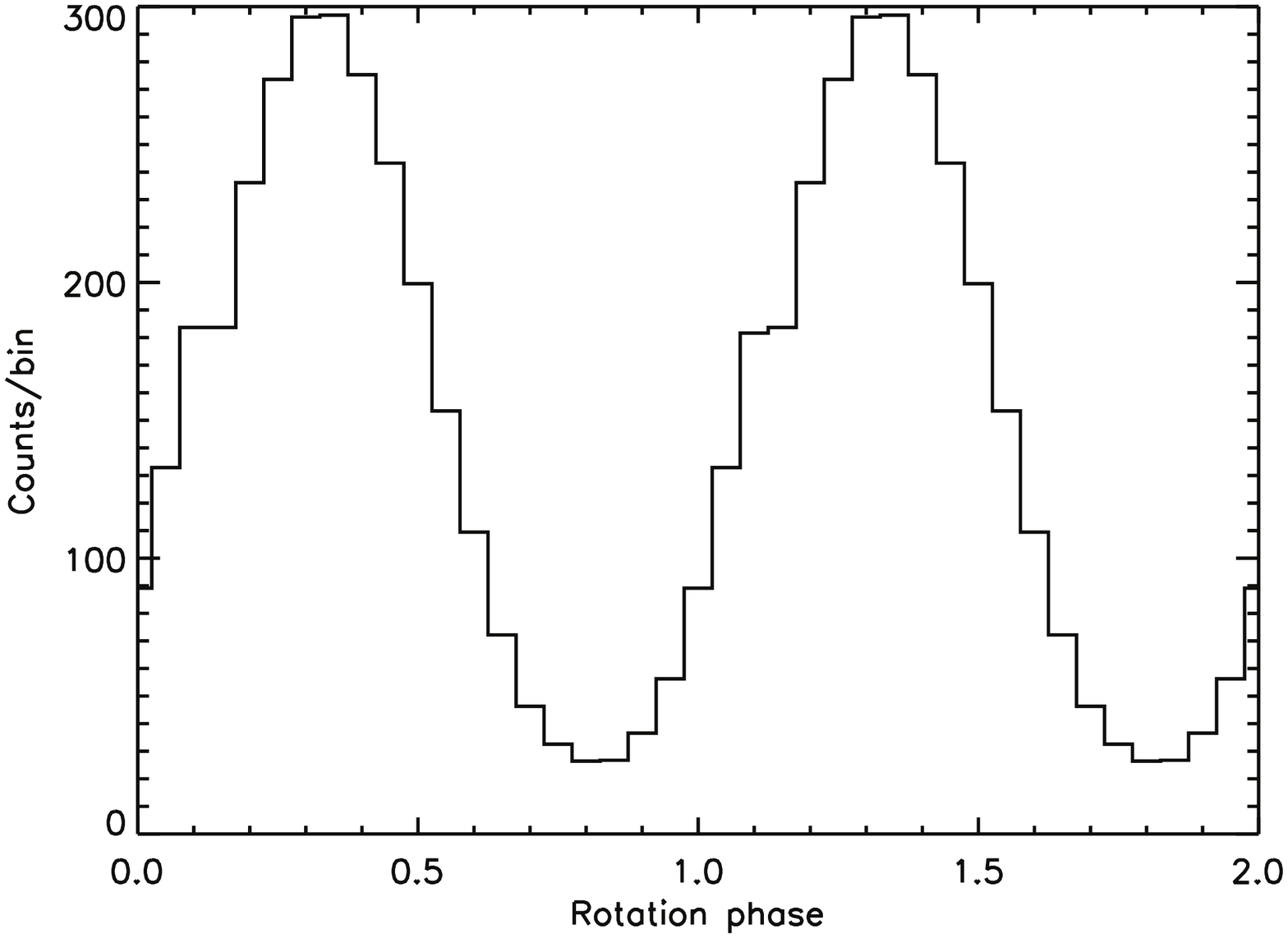}
\caption{{\it Top panel:} \cxo\/ pulse profiles ($2-10$\,keV) with 5
and 10 phase bins. 
 The shaded regions indicate the peak  interval (phases 0.15--0.35) used for phase-resolved spectroscopy. {\it Bottom panel:} 
 \cxo\/ pulse profiles ($2-10$ keV) with 10 phase bins, averaged over the reference phase. }
  \label{fig:pulse:acis:point}
\end{figure}

 Figure \ref{fig:pulse:acis:point} (upper panel) shows the pulse profiles with 5 and 10 phase bins.
 We used these profiles to measure the pulsed fraction\footnote{The pulsed fraction $p$ is defined as the ratio of the number of counts above the minimum level to the total number of counts.}, $p = 85\% \pm 10\%$.  We estimated the 
uncertainty of the pulsed fraction
using Monte Carlo simulations and bootstrapping, also accounting for the
time resolution and dead time in the 1/8 subarray mode. 
We also performed randomization of the arrival times within the 0.4\,s frame time and re-calculated the pulsed fraction, which remained within the uncertainty range estimated above. 

The pulsed fraction can also be defined as 
$\tilde{p}=[2(Z_n^2 - 2n)/N]^{1/2}$, where $n$ is the number
of harmonics that give a significant contribution,
 and $N$ is the number
of counts\footnote{The advantage of this definition is the independence of
$\tilde{p}$ of phase binning. For the case of purely sinusoidal  pulsations,
 $\tilde{p}$ coincides with $p$
 (assuming a very large number of bins and low noise),
while it is a factor of $\sqrt{2}$ larger than
 the RMS measure of variability. }.
 In our case,  $\tilde{p} = 1.13$ exceeds 100\%, which might be due to dead-time
effects and the relatively large ($\approx 0.18$) ratio of the 
time resolution to the period. To measure the pulsed fraction more accurately, the target should be observed with a better time resolution. 
 
Figure \ref{fig:pulse:acis:point} (lower panel) shows a 20-bin pulse profile 
 averaged over the reference phase\footnote{ This pulse profile was obtained by averaging 100 pulse profiles 
(20 bins each) constructed by assigning  different phases   
 to the first count and folding with the SGR period. }.
 We also produced a pulse profile for the surrounding extended emission  
 but did not find a statistically significant pulsed signal.

\section{Spectral Analysis}
\label {sec:spectra}

\subsection{{\it Swift}}
\label{sec:xrtspec}
We have fitted all XRT spectra ($2-10$\,keV) jointly with two continuum models: a power law (PL) and a single blackbody (BB), both with interstellar absorption. In the first case, we find that the photon index remains the same within the uncertainties;  therefore, we forced all observations in our joined fit to have the same varying photon index, while the normalizations were allowed to vary individually. We obtained a good fit ($\chi_{\nu}^2$ = 1.01 for 62 dof) with the best model parameters $\nh=10.5^{+1.9}_{-1.8}\times10^{22}$\,cm$^{-2}$, and photon index $\Gamma =3.2 \pm 0.4$. The absorbed BB model resulted in temperatures that also remained consistent within their uncertainties; we then linked the temperatures and allowed the normalizations to vary. We again obtained a good fit  ($\chi_{\nu}^2$ = 1.04 for 62 dof) with $ \nh=4.4^{+1.3}_{-1.2}\times10^{22}$\,cm$^{-2}$ and $kT = 1.1 \pm 0.1$\,keV. 
 The temperature is higher than those measured in most other magnetars \citep[typically around 0.5 keV; ][]{woods2006}.
 
\subsection{\cxo}
\subsubsection{CXOU\,J183452.1$-$084556}
 
We collected a total of 733 counts ($2$--$10$\,keV) from a circular region of $r=1''$ centered at CXOU~J$183452.1$--$084556$; the background contribution is expected to be only 0.25 counts (background was measured an $20''<r<33''$ annulus). We then grouped the source spectrum requiring a minimum of 15 counts per spectral bin. The resulting spectrum is shown in Figure~ \ref{fig:spec:acis:rxte} (black error bars). The source pileup is negligibly small ($\lesssim 1\%$), as the total source count rate of 0.057 counts\,s$^{-1}$ corresponds to 0.025 counts per frame. The point  source spectrum can be fitted equally well with  both the absorbed PL and BB models (see Table \ref{tab:cxofits}).  The observed (absorbed) flux ($2-10$\,keV) is $F_{\rm point}=( 3.0\pm0.1)\times10^{-12}$\,\cgs\  (corrected for the finite extraction aperture and $9\%$ deadtime). Table \ref{tab:cxofits} contains the values of the $\nh$ and photon index for the best-fit absorbed PL model, and the $\nh$ and temperature ($kT$) of the BB model. From the BB fit we estimate the emitting area radius to be $0.26$ km \citep[assuming that the source is at the same distance of 4\,kpc as the SNR W\,41;][]{tian2007}. The corresponding unabsorbed PL and BB fluxes  ($2-10$\,keV) are $1.6_{-0.4}^{+0.6}$ and $(5.8\pm0.6)\times10^{-12}$\,erg s$^{-1}$, respectively (see also Table \ref{tab:cxofits}). 

\begin{figure}
 \centering
\includegraphics[width=5.3in,angle=-90]{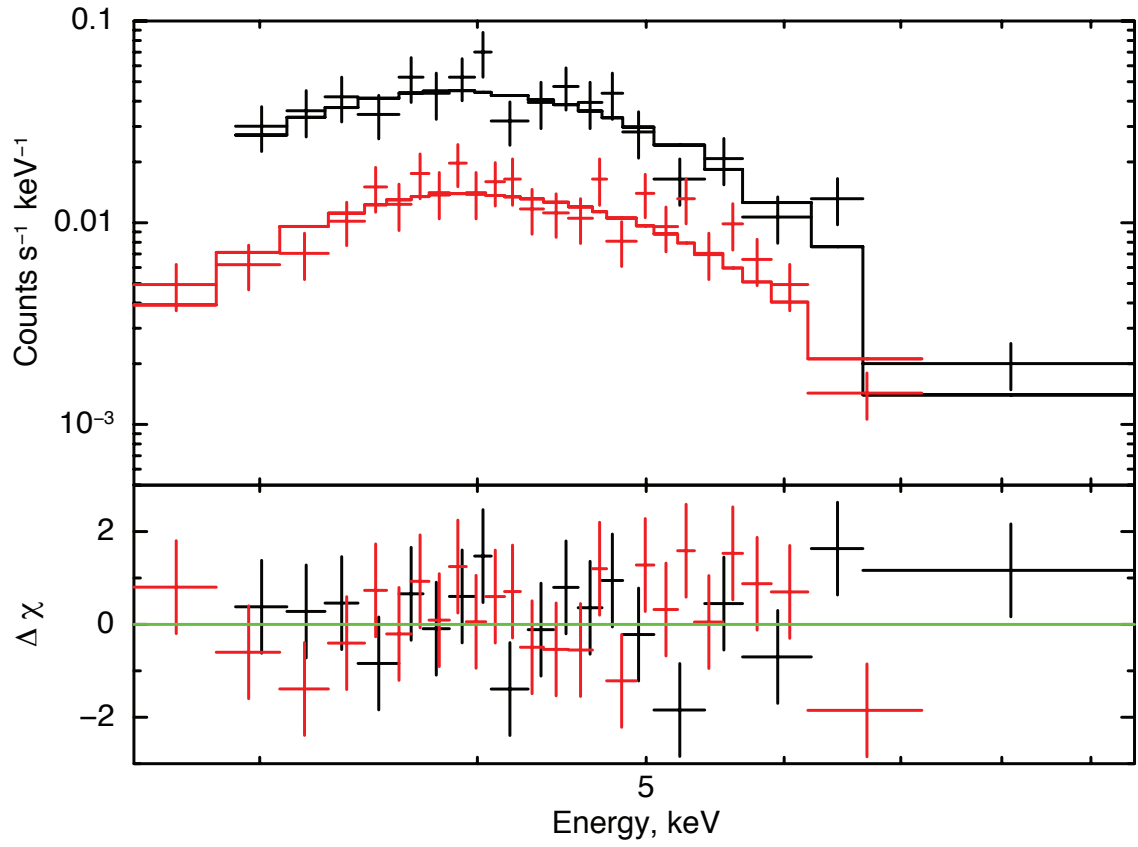}
\caption{PL fits of the \cxo\/ spectra of CXOU~183452.1--084556 in the pulse
maximum (black/bottom) and pulse minimum (red/top).
 }
\label{fig:spec:acis:phres}
\end{figure}

Guided by the pulse profiles shown in Figure \ref{fig:pulse:acis:point},  we have extracted spectra from two different phase intervals: 0.15--0.35 (peak) (indicated with the shaded region in Figure \ref{fig:pulse:acis:point} (upper panel)) and the rest (off-peak). These spectra are shown in Figure \ref{fig:spec:acis:phres}; we again used the PL and BB models in each case, fixing the $N_{\rm H}$ at the best-fit value of the phase-integrated spectrum. The peak and off-peak spectral parameters ($\Gamma$ or $kT$; see Table \ref{tab:phaseresspec}) are consistent within their uncertainties.

\subsubsection{Halo}
\begin{figure}
 \centering
\includegraphics[width=6.3in,angle=0]{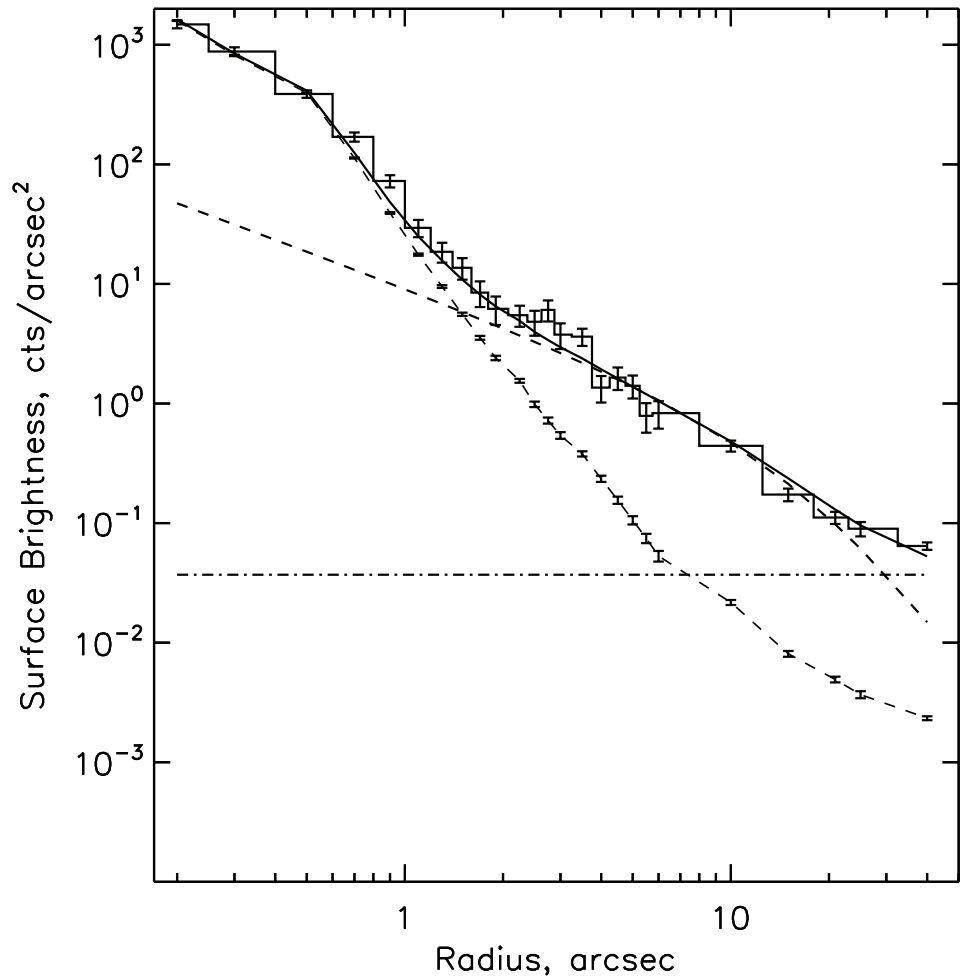}
\caption{Radial profile in $2-10$ keV from the 13 ks \cxo/ACIS observation (histogram) shown together with the simulated PSF (dashed line with error bars), a dust halo model (dashed line) from Misanovic et al.\  (2011), and a background (horizontal dash-dotted line) measured from the current observation. The solid line shows the sum of all three components.}
\label{fig:simulation}
\end{figure}

\begin{figure}
 \centering
\includegraphics[width=5in,angle=90]{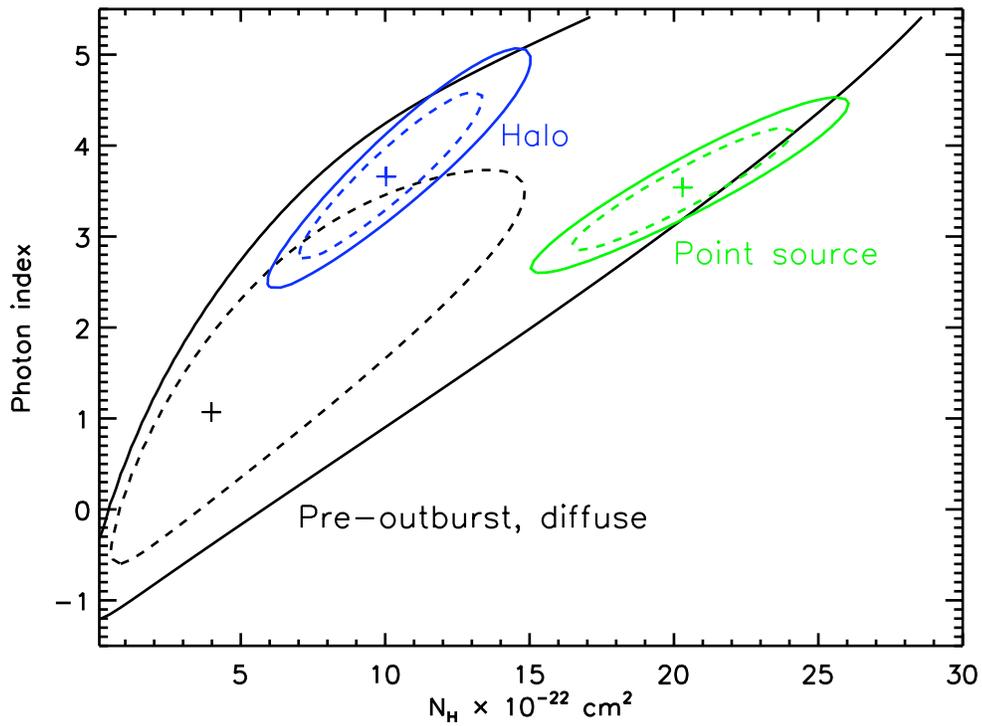}
\caption{ Confidence contours (68\% and 90\%) in the $N_{\rm
H}$--$\Gamma$ plane for the PL fit to the halo  ({\em blue}),
point source 
({\em green}),
 and the pre-outburst diffuse emission ({\em black}) spectra. The contours
 are obtained with the PL normalization fitted at each point
of the grid.  The best-fit parameter values are shown by crosses. 
}\label{fig:contours}
\end{figure}

We collected 314 counts ($2$--$10$\,keV) from the $2''<r<10''$ annulus (hereafter ``halo''), centered at CXOU\,J183452.1$-$084556,
 where the contribution of the point source is expected to be small ($<10\%$). We subtracted the background (estimated from a much larger region away from the source) and obtained a net total of $\simeq 300$\,counts. To separate the halo from the point source, we simulated a point spread function (PSF) using MARX\footnote{\url{http://space.mit.edu/cxc/marx/}}. The comparison of the data with the PSF simulation (Figure \ref{fig:simulation}) shows a good agreement within a small aperture (approximately up to 1$''$ radius), while the extended emission dominates at larger radii. Based on our simulation, we estimate that  $\sim33$\,photons come from the point source, after taking into account the extended PSF wings. The final halo spectrum was also binned requiring a minimum of 15 counts per bin. The best-fit PL slope is approximately the same as that of the point source
spectrum, while the best-fit $\nh$  is a factor of two lower 
 (see Table \ref{tab:cxofits}). The $\nh$-$\Gamma$ confidence contours for the
halo spectrum, together with those for the point source spectrum, are shown in
Figure \ref{fig:contours}.  The absorbed and unabsorbed  fluxes ($2$--$10$ keV)
of the halo emission are  $F_{\rm halo}=(4.7\pm0.2)\times10^{-13}$  and $(1.6\pm0.4)\times10^{-12}$\,\cgs, respectively 

The extended emission is well described by the same dust halo model as the one used by \citet{misanovic2011} for another nearby source, 
CXOU~J183434.9$-$084443,  according to which most of the dust must be located relatively close to the source (within 1/4 of the distance). At least part of this dust could be associated with the molecular cloud that appears to be interacting with W41 \citep{leahy2008}, in agreement with the very large  absorption column that we find in our spectral analysis (see Table 2).

\subsubsection{Pre-outburst \cxo/ACIS data taken on June 2009}
\label {sec:cxo-old}

\begin{figure}
\vspace{-0.8cm}
 \centering
\includegraphics[width=5.5in,angle=0]{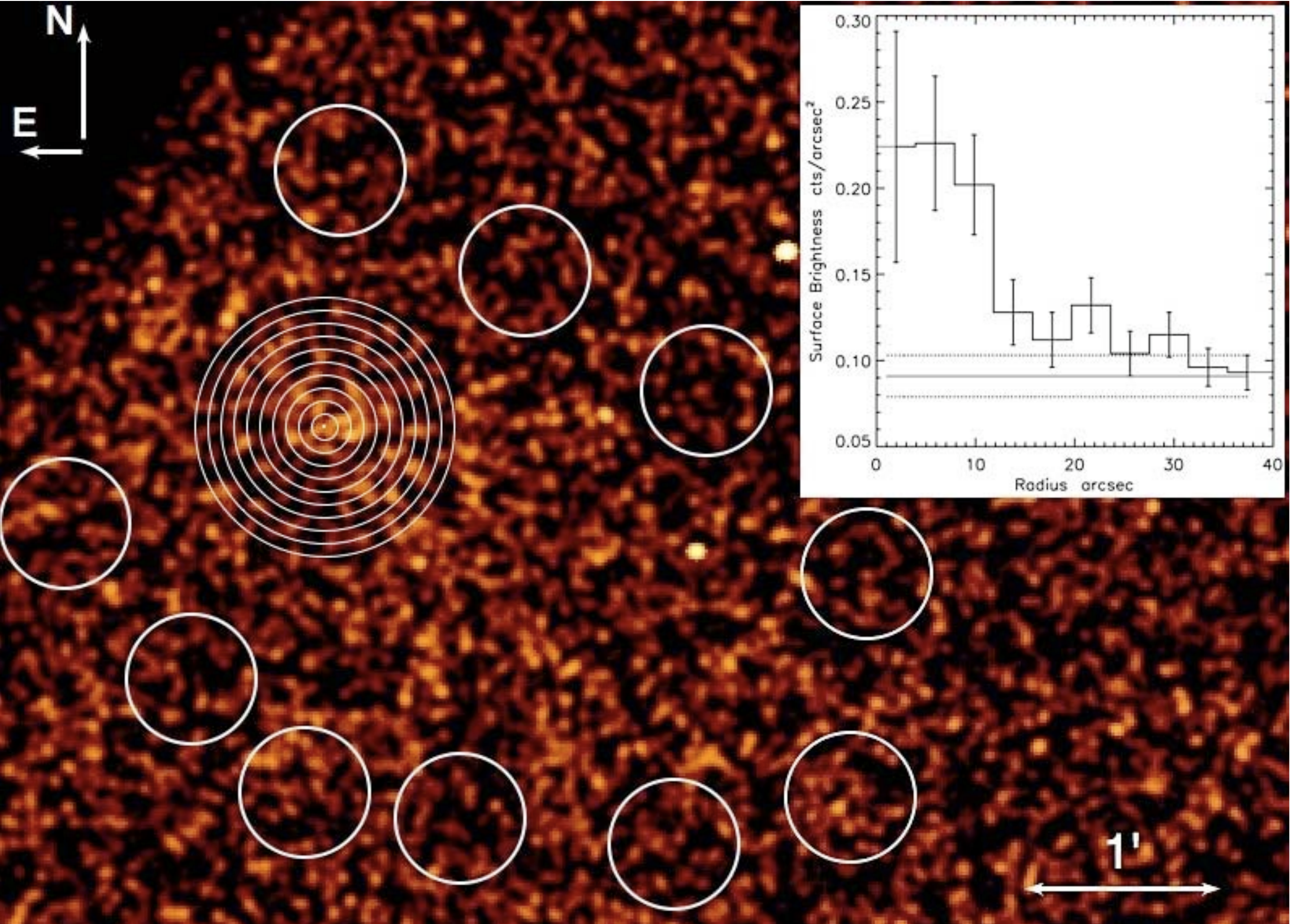}
\includegraphics[width=5.5in,angle=0]{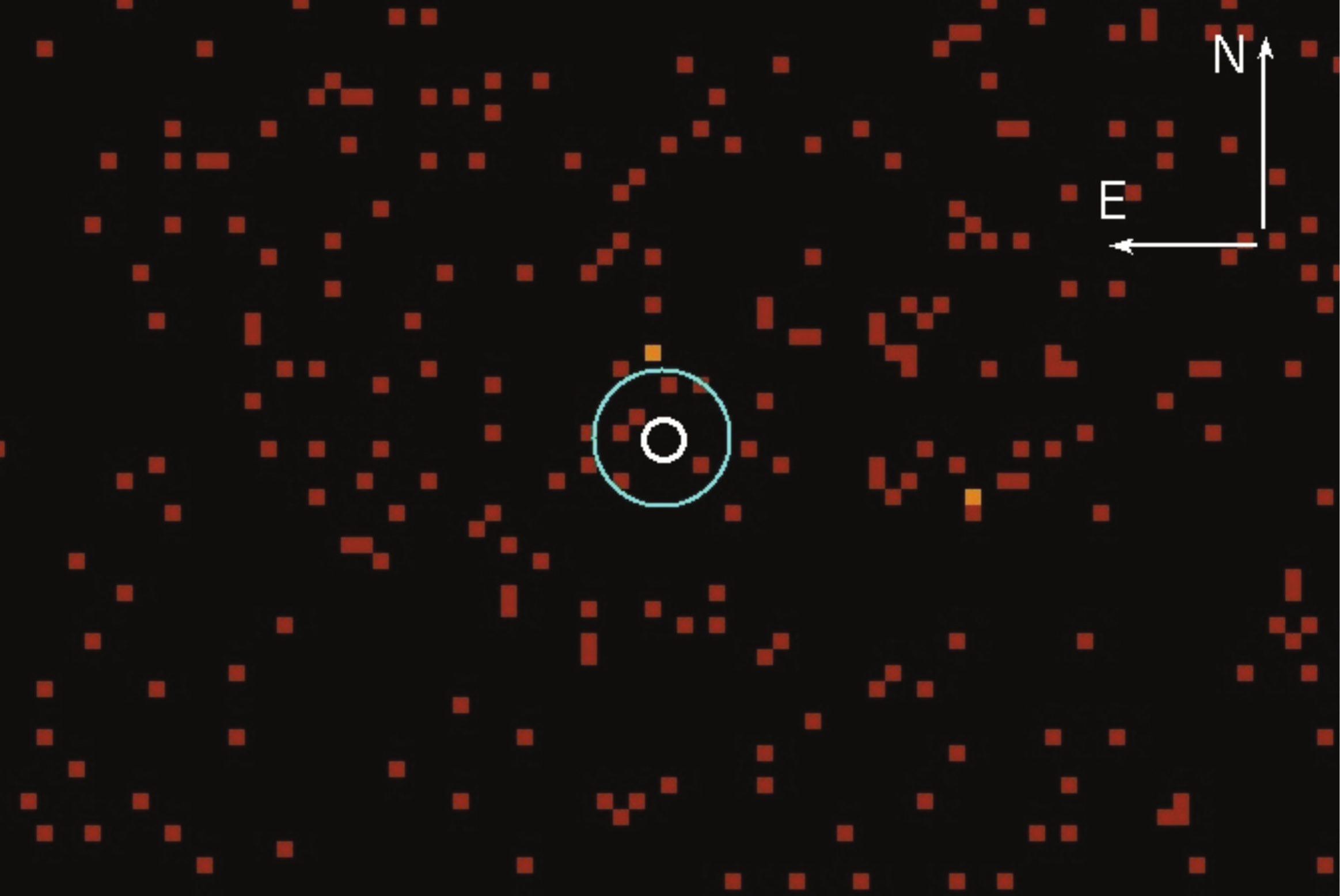}
\caption{Archival 2-8 keV  ACIS-S3 images of  the CXOU~J183452.1--084556 field obtained on 2009  June 7. {\it Upper panel:} 
Binned (pixel size $0\farcs98$) and smoothed (with a $2\farcs9$ Gaussian kernel) image of the field. The annuli, centered at the SGR 
 position, are used to extract the radial profile of the surface brightness distribution (shown in the inset). The background (solid horizontal line in the inset) and its uncertainty (dotted horizontal lines) are measured from the ten $r=20''$ circular regions.  
{\it Lower panel:} Zoomed-in image of the field around the position of CXOU~J183452.1--084556 at the native ACIS-S3 binning with no smoothing applied. The circles with radii $0\farcs6$ and $2''$ were used to estimate the detection upper limit (see Section \ref{sec:cxo-old}).}
\label{fig:img:acis:old}
\end{figure}

We analyzed the 2009 \cxo\/ observation covering the \sgr field and found zero photons within   the error circle ($r=0\farcs 6$) of CXOU~183452.1$-$084556 (see Figure \ref{fig:img:acis:old}). The off-axis angle of $\approx4\farcm 6$ during that observation is, however, large enough for the angular resolution to be substantially degraded compared to on-axis. Hence, to estimate the 2009 upper limit on the source flux, we used a larger radius, $r=2''$, which would contain about $50\%$ of the flux of a point source at this off-axis angle.  We found 5 and 4 photons in the $0.5$--$10$ and $2$--$10$\,keV bands, respectively. The mean local background surface brightness is $0.24\pm0.02$ and $0.18\pm0.01$ counts arcsec$^{-2}$ in $0.5$--$10$ and $2$--$10$\,keV, respectively. Thus, within the $r=2''$ extraction aperture we would expect to detect about two counts from the background. This translates into an upper limit of $~0.15$ counts ks$^{-1}$ in $2-10$\,keV, which corresponds to an absorbed flux limit of  $F_{\rm point}<(2-4)\times10^{-15}$\,\cgs.

 Although we do not detect a point source,  we notice extended emission on larger scales around the position of \sgrnos. We demonstrate this by plotting the radial profile of the surface brightness (see Figure \ref{fig:img:acis:old}, top panel). The ten annuli used to extract the radial profile are centered at the position of \sgr,  while the background is measured from ten circular ($r=20''$) regions surrounding the source. One can see from the figure  inset 
  that most of the excess over the background is within $r\lesssim12''$ and it corresponds to a detection significance of $\approx5.1\sigma$. However, there is also  marginal ($\approx3\sigma$) evidence for extended emission at larger scales  (between $r=12''$ and $30''$, see Figure \ref{fig:img:acis:old}, top). 
 There might be an even more extended (primarily toward southwest from the \sgr), fainter asymmetric emission, but its significance can only be established 
 with deeper \cxo/ACIS observations.

We also found evidence for extended emission  in the 2005  \xmm\/ data \citep[see also][]{mukherjee2009}. The extent, location  and significance of the
 extended emission in the   EPIC/MOS images (which are not affected by the chip gaps and have a  low enough background) are similar to those measured from the 2009  \cxo/ACIS images (see above). The previously reported large-scale extended emission west-northwest of the SGR (i.e., in the direction toward CXOU183434.9--084443) could be mainly due to the point sources that are smeared out in the EPIC images because of  the coarse angular resolution of \xmm. The two brightest point sources are clearly resolved in the sharper  \cxo/ACIS images (cf.\ corresponding panels in Fig.~\ref{fig:img:acis:old}). 
 We did not attempt to extract the spectra from the 2005 \xmm/EPIC data because the background is much higher  and the angular resolution is worse than the one of the \cxo\/ observation. No point source is detected in the \xmm\/ images at the position of CXOU~J183452.1$-$084556.  We have not estimated an \xmm\/ upper limit on the point source flux, as it would be less restrictive than the one derived using the  2009 \cxo/ACIS data.

\subsection{Joint fits of \cxo\/ and \rxte\/ data}
\label {sec:cxo-rxte}

Since the \rxte/PCA is not an imaging instrument, we could not spatially separate the halo and the point source or even subtract a background measured independently from an offset region.  However, since the instrument has a broader spectral range than the {\sl Swift}/XRT and \cxo/ACIS, potentially providing valuable source information above 10 keV, we used the latter data to calibrate our PCA  spectrum of the longest (9.7\,ks) pointing of 2011 August 9 (see Table \ref{tab:xraylist}). For the \rxte\/ data, we accumulated the spectrum at the pulse minimum (which contains background, halo, and any unpulsed point source contributions) as the background and subtracted it from the source spectrum integrated over the remaining phases (see shaded regions in Figure \ref{fig:pulse:rxte}). The resulting pulsed emission spectrum was then rebinned to have at least 50 counts per spectral bin after the background subtraction.

\begin{figure}
 \centering
\includegraphics[width=5.3in,angle=0]{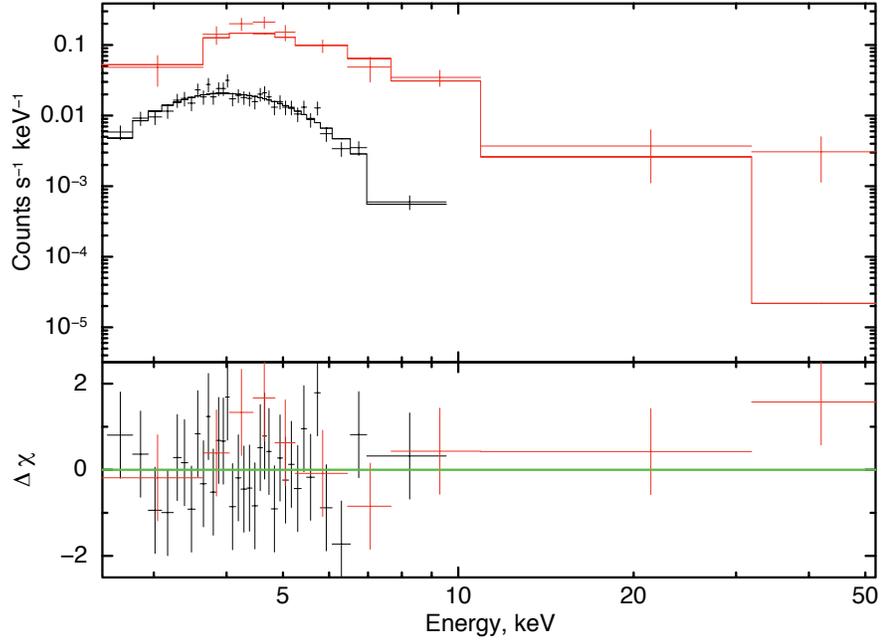}
\includegraphics[width=5.3in,angle=0]{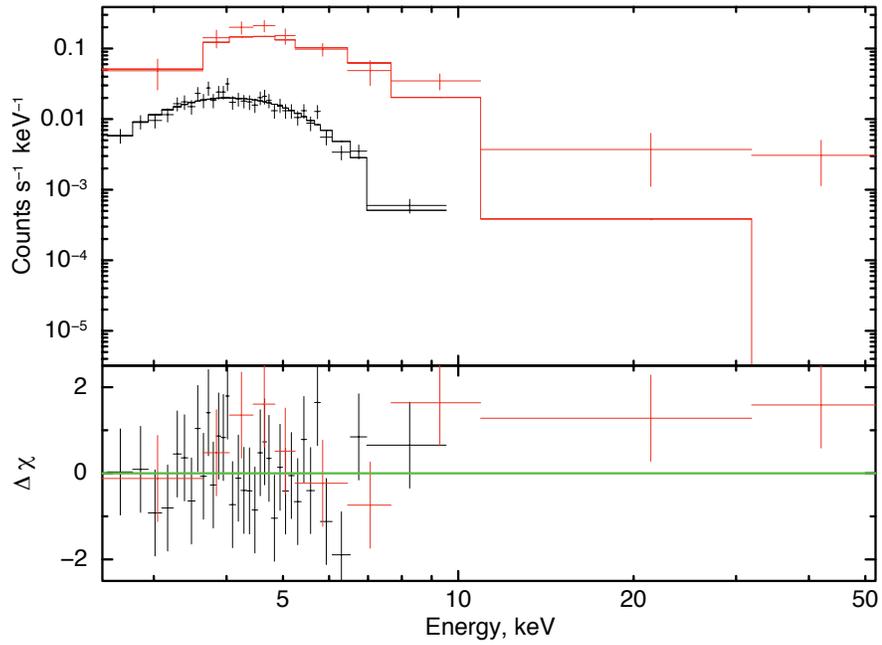}
\caption{
 \cxo/ACIS (black/bottom) and \rxte/PCA  (red/top) spectra
of  CXOU~183452.1$-$084556 jointly fitted  with the  PL and BB models
(top and bottom panels, respectively).}
\label{fig:spec:acis:rxte}
\end{figure}

We performed a joined spectral fit of the \rxte\/ ($2$--$50$\,keV)  and the \cxo/ACIS  ($2$--$10$\,keV) data
 (see Figure \ref{fig:spec:acis:rxte}). We found that the best fits are obtained when the \rxte\/ flux is scaled down by a factor $\eta=0.6$. The resulting PL best-fit parameters are very close to those of the \cxo/ACIS fits but somewhat better constrained (see Table \ref{tab:cxofits}). A single BB fit is disfavored by systematic residuals at energies $>8$\,keV  (see Figure \ref{fig:spec:acis:rxte}, bottom panel).
 The introduced scaling of the \rxte\/ flux can be interpreted as due to two reasons: (1) the source was brighter at the time of the \rxte\/ observation, and (2) the true background is lower than that estimated from the pulse minimum (see above).

\section{Discussion}
\label{sec:discussion}
\subsection{\sgrnos}

\sgr has one of the shortest periods among magnetars\footnote{See the McGill AXP/SGR catalog: \url{http://www.physics.mcgill.ca/$\sim$pulsar/magnetar/main.html}.}
and one of the highest pulsed fractions of the persistent X-ray emission, 
similar to that of 1E 1048.1$-$5937 on 2000 December 28 \citep{tiengo2002}.  In its timing and spectral properties, \sgr\ strongly resembles
the recently discovered  SGR\,J1833$-$0832, which has a period of 7.6 s and a magnetic field of $1.8\times 10^{14}$ G  \citep{gogus2010,esposito2011}.
 In particular,
similar to SGR\,J1833$-$0832 (and unlike most other SGRs with good quality spectra), the $0.5-10$\,keV spectrum of \sgr can be fitted with a single BB model,
 whose temperature, $kT\simeq 1.1$ keV, is the same as that of SGR\,J1833$-$0832.
 The \sgr BB radius, $R=0.26$ km,
is a factor of 3 smaller than in SGR\,J1833$-$0832,
 which, however, may not be a significant difference given the poorly known distances. Another similarity between  \sgr and SGR\,J1833$-$0832 is the lack 
of obvious spectral shape evolution with rotational phase. The phase-resolved spectra (see Figure \ref{fig:spec:acis:phres})  differ only in normalization, and the  differences in other model parameters are not statistically 
significant. 
 Despite these 
similarities, the post-burst flux decay trend is markedly different for the 
two SGRs.  The unabsorbed flux of \sgr decreased as $\propto t^{-0.53\pm0.07}$ from day 2 after the burst  (Figure \ref{fig:lcurve}),
while the flux of SGR\,J1833$-$0832 remained constant for nearly 20 days before the onset of decline. We note, however, that this early constancy of the flux in SGR\,J1833$-$0832 is unusual; the enhanced persistent X-ray flux of magnetars following an outburst usually declines 
 as a power law with an index similar to that of \sgrnos. 

Similar BB temperatures and radii were also found for SGR~0418+5729  \citep[$P=9.1$ s, $B<7.5\times 10^{12}$ G;][]{esposito2010}
 from the 
 {\sl Swift} XRT data  taken within $\sim 10$ days after the outburst. Also, SGR~0418+5729 exhibited a $\propto t^{-0.3}$ decay  during the first 19  days and a much steeper, $\propto t^{-1.2}$, decay thereafter. 

It is tempting to interpret the small emitting area of  \sgr
(similar to those of SGR\,J1833$-$0832 and SGR~0418+5729) as a hot spot on the neutron star surface. 
  We should note that it would be very difficult to obtain such a high pulsed fraction even for a very small hot spot emitting (nearly isotropic) BB radiation because the pulsations would be washed out by the light bending in the 
neutron star gravitational field \citep[see, e.g.,][]{zavlin1995}. If, however, we take into account that the angular distribution of radiation from a
neutron star atmosphere has a narrow peak along the magnetic field direction 
\citep{pavlov1994}, such a high pulsed fraction can indeed be explained assuming that the
observed radiation emerges from a small hot spot near the magnetic pole
of the neutron star. The fact that the light curve with such a high pulsed
fraction shows only one peak per period suggests that the magnetic field
configuration is substantially different from a centered dipole (e.g.,
it could be a strongly decentered dipole, in which case the magnetic
fields and the temperatures are substantially different at the two poles).  
We caution here that 
the BB model provides only an empirical description of the spectral shape. 
It can be used for comparison of different sources,
 but it may be significantly different from the actual spectrum
emitted from a neutron star atmosphere
\citep{pavlov1995} and possibly modified by the resonance Compton scattering in the neutron star magnetosphere \citep{nobili2008}.
 Comparing the BB fit parameters of the three SGRs (J1834.9$-$0846, J1833$-$0832, and J0418$+$5729), we can conclude that they do not depend on the SGR period
(in the range of 2--10\,s), nor on the strength of the 
 spin-down magnetic field (in the range of (0.1--$2)\times 10^{14}$\,G).

    The nondetection of the \sgr in the pre-outburst  \cxo\/ data shows that the SGR flux can vary by at least a factor of $\sim 10^3$
 between the presumably truly quiescent level in the low state
 and an elevated level 
 that 
 has persisted, with a slow decay, for at least 6 weeks after  the outburst. 
   This suggests that there is  a large number of SGRs in a quiescent
state undetectable at the current level of sensitivity of X-ray observatories.

    \subsection{Extended emission}
    
  At first glance,
 the extended emission around \sgrnos, detected by {\sl CXO} in 2011, looks rather unusual.
Its radial distribution is consistent with that of a dust scattering halo (see 
Figure \ref{fig:simulation}), but its 
 spectrum shows some peculiarities.
In particular, the 
 best-fit hydrogen column density of the halo is 
a factor of 2 lower than that of the central source, while their 
spectral slopes are similar within statistics, instead of being steeper by $\Delta\Gamma =1$--2, as
expected for the model halo spectrum \citep[see, e.g.,][]{misanovic2011}.
A likely explanation can be derived from the breadth of the 
$\nh$--$\Gamma$ confidence contours and the strong correlation of these parameters.
Indeed, Figure \ref{fig:contours} 
shows that an intermediate $\nh \approx 1.5\times 10^{23}$\,cm$^{-2}$ corresponds to the 90\%  point source and halo confidence contours, and
the best-fit photon indices at such $\nh$ are $\Gamma\approx 3$ and $\approx5$ for the
point source and halo, respectively. Thus, we believe that the dust scattering
halo is the most plausible interpretation of the extended emission around \sgrnos.

We should also note that a fainter extended emission was seen around the
magnetar positition in the archival \cxo\/
 data from 2009, in which no point source was detected. Although the best-fit $\nh$ and $\Gamma$ for
this pre-outburst emission are substantially smaller than those in the
post-outburst data (see Table \ref{tab:cxofits}), the large uncertainties of these parameters make them consistent with the corresponding parameters measured for the halo in 2011.   The existence of a halo in the archival data may indicate that \sgr experienced an outburst not long before the 2009 June 7 observation.

Although it seems certain that most of the extended emission is the dust-scattered emission from the magnetar,
 we cannot exclude the possibility that it may 
contain some kind of a pulsar wind nebula (PWN), due to synchrotron
radiation from relativistic electrons/positrons accelerated in the neutron star 
magnetosphere and shocked in the ambient medium. We know that in the case of
rotation-powered pulsars, a typical X-ray PWN luminosity is about
$10^{-4} \dot{E}$, albeit with a large scatter \citep{kargaltsev2008}. If the same relationship
is valid for magnetars, we would expect $L_{\rm pwn} \sim 10^{30}$\,erg~s$^{-1}$,
  which would be undectable at the presumed distance of 4\,kpc. It might happen, however, that a ``magnetar wind nebula'' is more efficient than one created
by a rotation-powered pulsar, in which case we would expect a detectable contribution. To separate it from the dust scattering halo, one
should analyze several data sets obtained at different times after the 
outburst. We expect that the halo component flux would be changing in proportion to the point
source flux (with a time lag), while the PWN component would remain constant.

 \subsection{Relation to SNR\,W41 and HESS \,J$1834$--$087$}
 
     The distance to \sgr still remains an open issue. 
 As this source is located within W41, association with this SNR is certainly plausible \citep[other SGRs were found near SNR centers,][]{woods2006}, but it 
 has not been firmly proven.
    Similar extreme absorption \citep[$\nh=3\times 10^{23}$\,cm$^{-2}$;][]{misanovic2011} 
has been measured for the neighboring  CXOU~J183434.9$-$084443,
  indicating that such an absorption is not a unique feature 
of \sgr and hence not intrinsic to it. 
However, the distance to (and the origin of) CXOU~J183434.9$-$084443 are also 
unknown. It could be a pulsar associated with W41 or a background AGN located much farther. \cite{leahy2008} presented evidence for molecular clouds near W41,
 which are likely interacting with the SNR. The large absorbing column could be attributed to those clouds. At this point we can only conclude with certainty that \sgr is at the distance of 4\,kpc or farther (the line of sight 
in that direction intersects several spiral arms). 
To better constrain the distance, the method of \citet{durant2006} could be used; however, it requires grating observations, which are only feasible when the source is in the bright state.  We note, however, that the conclusion by \citet{durant2006}, that  all AXPs have more or less standard luminosity of $1.3\times 10^{35}$\,erg~s$^{-1}$, cannot  hold for quiescent SGRs because otherwise 
 they would have been easily seen even at the most extreme distance of 20\,kpc.

\begin{figure}
 \centering
\includegraphics[width=7.1in,angle=0]{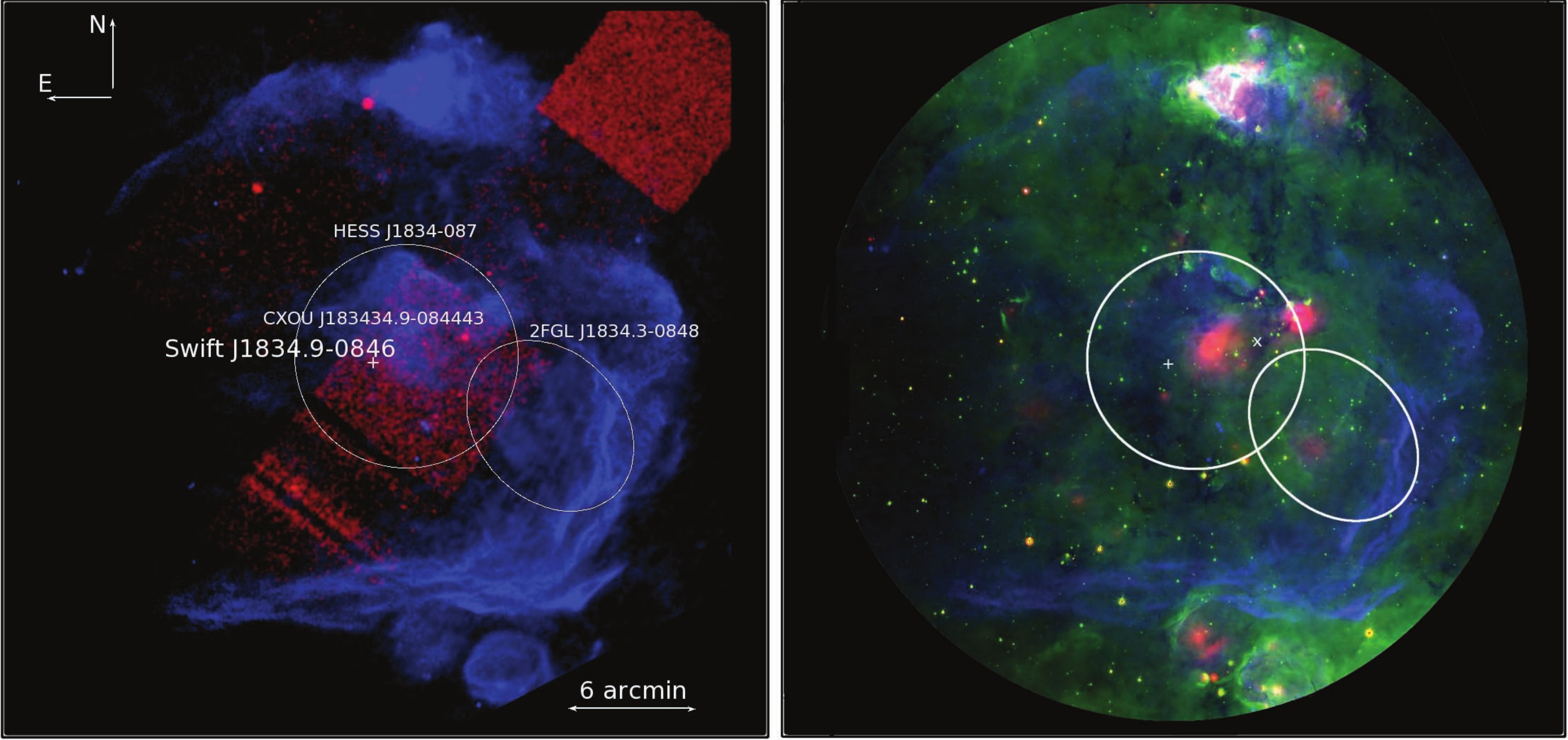}
\caption{Multiwavelength emission from  the W41 region. The left panel shows 
the \cxo\/ ACIS image (0.3--8\,keV; red) and the VLA 20 cm image (blue) from 
the MAGPIS database (http://third.ucllnl.org/gps/).   
In the right panel 
the 20 cm VLA image is shown in blue, the 
{\sl Spitzer} IRAC 8 $\mu$m image in green, and the {\sl Spitzer}  MIPS 24 $\mu$m image
in red.
 The  $r=5\farcm4$ circle shows the extent of HESS\,J1834--087 \citep{aharonian2005}; the ellipse shows the position of
2FGL\,J1834.3--0848 at the 95\% confidence level
 \citep{abdo2011}. }
\label{fig:img:multiwave}
\end{figure}

The field surrounding \sgr is rich with high-energy sources (see Figure \ref{fig:img:multiwave}). This magnetar
is located at the heart  of SNR\,W41 and nearly at the center  of the  extended TeV source  HESS\,J1834$-$087, which itself is confined to the SNR interior (see Figure \ref{fig:img:multiwave}). In addition, there is a somewhat offset {\sl Fermi} source, 2FGL\,J1834.3$-$0848, located nearby
 (see Figure \ref{fig:img:multiwave}). Since the extent of HESS\,J1834$-$087 is significantly smaller than that of the SNR, the TeV emission cannot be coming from the SNR shell as it does in some other cases \citep{bochow2011}. The only other plausible explanation is that the TeV emission is powered by relativistic electrons injected by the compact object formed after the SNR explosion. There are currently several candidates for such an object. Firstly, a few SGRs are known to be associated with shell-type SNRs \citep{hurley2000}, and the central location of \sgr  certainly supports such a hypothesis.  On the other hand, there is no firm evidence so far that
 SGRs can produce copious amounts of relativistic particles similar to 
young rotation-powered pulsars. While there is a convincing evidence that pulsars can power relic PWNe
 emitting TeV $\gamma$-rays, such evidence is currently lacking for magnetars.
 
\begin{figure}
 \centering
\includegraphics[width=6.8in,angle=0]{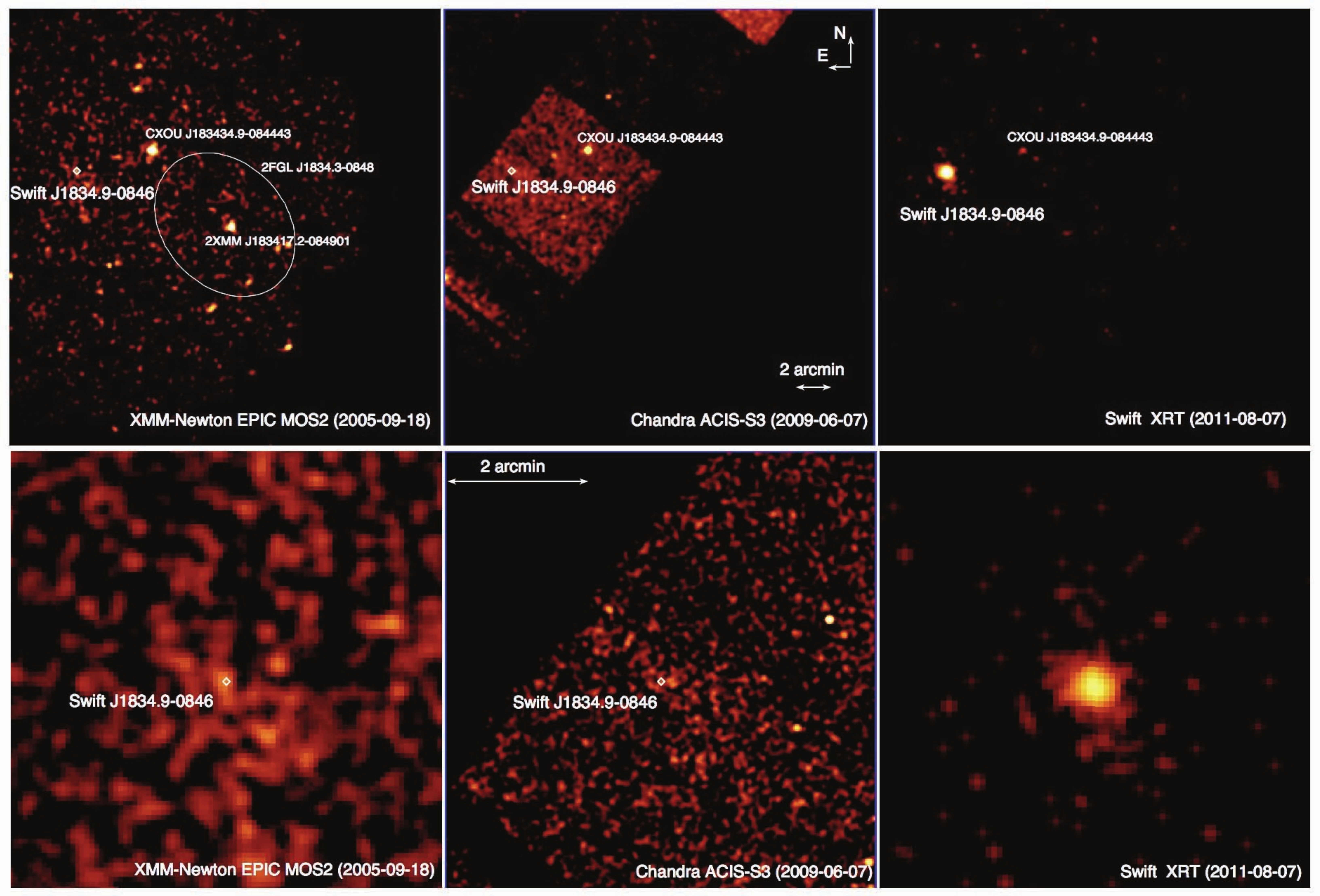}
\caption{Mosaic of 
images (2-10 keV) of the central region
 of the W41 SNR, spanning six years. 
The left, middle, and right columns  correspond to 
the data obtained with \xmm/EPIC (2005 Sept 18), \cxo/ACIS-S (2009 June 7), and {\sl Swift}/XRT (2011 August 7), respectively. The lower panels are  zoomed in the \sgr position.
}
\label{fig:img:acis:xmm:swift}
\end{figure}

Among other sources possibly related to HESS\,J1834$-$087 and W41 are CXOU J$183434.9-084443$ \citep[a PWN candidate discussed in detail by][]{misanovic2011} and the 2XMM J183417.2$-$084901, which is located right at the center of the {\sl Fermi} error circle (see Figure~\ref{fig:img:acis:xmm:swift}, top panels). Further longer observations of this region are required to understand the connection between the sources observed in different energy domains.  
 
\acknowledgments
 
The authors are grateful to Harvey Tananbaum for his decision to award his DDT time for \cxo\/ observations of \sgrnos. SW would like to thank Davy Kirkpatrick for the use of his Palomar observing time to obtain the near-IR observations of \sgrnos. This work was partly based on observations obtained at the Hale Telescope, Palomar Observatory, as a part a continuing collaboration between the California Institute of Technology, NASA/JPL, and Cornell University.
This publication makes use of data products from the 2 Micron All Sky Survey, which is a joint project of the University of Massachusetts and the Infrared Processing and Analysis Center/California Institute of Technology, funded by the National Aeronautics and Space Administration and the National Science Foundation. 
The work by OYK and GGP was partly supported by NASA grants NNX09AC81G and
NNX09AC84G, NSF grants AST09-08733 and AST09-08611, and by the Ministry of Education and Science of the Russian Federation (contract 11.G34.31.0001). CK was partly supported by NASA grant NNH07ZDA001-GLAST.
LL is supported by the Postdoctoral Research Program of the Turkish Academy of Sciences (T\"UBA).

\begin{deluxetable}{lllll}
\tabletypesize{\scriptsize} \tablewidth{0pt}
\setlength{\tabcolsep}{0.1in} \tablecaption{X-ray observations of  Swift J1834.9-0846.\label{tab:xraylist}} 
\tablehead{\colhead{Date} & \colhead{ObsID}& \colhead{Observatory/Detector (Mode) } & \colhead{Exposure, ks} &  \colhead{Time resolution, s }} 
\startdata 
2005 Sept 18 & 0302560301 & \xmm\/ EPIC & 18.6 & 0.072 \\
2009 Jun 7 & 10126 & \cxo\/ ACIS-S & 46.5 &  3.2 \\
2011 Aug 7 & 00458907000 & {\it Swift}/XRT (PC) & 1.54 & 2.5 \\
2011 Aug 7\tablenotemark{a} & 00458907001 & {\it Swift}/XRT (WT) & 0.096 & $1.8\times10^{-3}$ \\
2011 Aug 8\tablenotemark{a} & 00458907002 & {\it Swift}/XRT (WT) & 0.129 & $1.8\times10^{-3}$ \\
2011 Aug 8 & 00458907003 & {\it Swift}/XRT (WT) & 1.65 & $1.8\times10^{-3}$ \\
2011 Aug 8 & 00458907004 & {\it Swift}/XRT (WT) & 0.958 & $1.8\times10^{-3}$ \\ 
2011 Aug 9 & 00458907006 & {\it Swift}/XRT (WT) & 2.67 & $1.8\times10^{-3}$ \\
2011 Aug 9 & 96434-01-01-00   &   \rxte\ PCA  & 3.40  & $9\times10^{-7}$  \\
2011 Aug 9 & 96434-01-02-00  &   \rxte\ PCA  & 9.66 & $9\times10^{-7}$  \\
2011 Aug 12 & 00458907007 & {\it Swift}/XRT (WT) & 5.67 & $1.8\times10^{-3}$ \\
2011 Aug 14& 00458907008 & {\it Swift}/XRT (WT) & 5.39 & $1.8\times10^{-3}$ \\ 
2011 Aug 14 & 96434-01-03-00 &  \rxte\ PCA  &  6.78  & $9\times10^{-7}$ \\
2011 Aug 18  & 96434-01-03-01  &  \rxte\ PCA  & 6.75  & $9\times10^{-7}$ \\
2011 Aug 18& 00458907009 & {\it Swift}/XRT (WT) & 5.73 & $1.8\times10^{-3}$ \\
2011 Aug 21 & 00458907010 & {\it Swift}/XRT (WT) & 2.49 & $1.8\times10^{-3}$ \\
2011 Aug  22 & 14329 &   \cxo\/ ACIS-S & 13.0 & 0.44104 \\
2011 Aug 24 & 96434-01-04-00  &  \rxte\ PCA  & 6.60  & $9\times10^{-7}$ \\
2011 Aug 24\tablenotemark{a} & 00458907011 & {\it Swift}/XRT (WT) & 0.94 & $1.8\times10^{-3}$ \\
2011 Aug 27 & 00458907012 & {\it Swift}/XRT (WT) & 1.95 & $1.8\times10^{-3}$ \\
2011 Aug 29 & 96434-01-05-00 &    \rxte\ PCA  & 6.05  & $9\times10^{-7}$ \\
2011 Aug 30 & 00458907013 & {\it Swift}/XRT (WT) & 2.16 & $1.8\times10^{-3}$ \\
2011 Sep 2 & 96434-01-06-00   &  \rxte\ PCA  & 5.12  & $9\times10^{-7}$ \\
2011 Sep 2\tablenotemark{a} & 00458907014 & {\it Swift}/XRT (PC) & 2.06 & 2.5 \\
2011 Sep 5\tablenotemark{a} & 00458907015 & {\it Swift}/XRT (PC) & 1.72 & 2.5 \\
2011 Sep 8 & 96434-01-06-01    &  \rxte\ PCA  & 5.52  & $9\times10^{-7}$ \\
2011 Sep 10\tablenotemark{a} & 00458907016 & {\it Swift}/XRT (PC) & 2.01 & 2.5 \\
2011 Sep 15 & 00032097001 & {\it Swift}/XRT (WT) & 9.09 & $1.8\times10^{-3}$ \\
2011 Sep 18 & 00032097002 & {\it Swift}/XRT (WT) & 10.45 & $1.8\times10^{-3}$ \\
2011 Sep 21 & 00032097003 & {\it Swift}/XRT (WT) & 7.44 & $1.8\times10^{-3}$ \\
2011 Sep 24 & 00032097004 & {\it Swift}/XRT (WT) & 8.10 & $1.8\times10^{-3}$ 
\enddata
\tablecomments{  {  Log of all observations used in our analysis. }}
 \tablenotetext{a}{Excluded from the spectral analysis.}
\end{deluxetable}

 \clearpage
 
 \begin{table}[]
\caption[]{Spectral fit parameters of \sgr (point source) and its extended emission (halo).\label{tab:cxofits}} \vspace{-0.5cm}
\begin{center}
\begin{tabular}{ccccccccc}
\hline 
Model & $N_{\rm H,22}$   & $\mathcal{N}$\tablenotemark{a}  or    $R$\tablenotemark{b} & $\Gamma$\tablenotemark{c}  or  $kT$\tablenotemark{c} &  $\chi_{\nu}^2$/d.o.f.\tablenotemark{d}   & $L_{\rm X}$\tablenotemark{e}  \\
\hline
CXO, PL (point) & $20.3^{+2.7}_{-2.5}$ &
$4.8_{-2.4}^{+5.9}$ & $3.5_{-0.4}^{+0.5}$ & $0.72/31$ &  
3.1 \\
CXO, BB (point) & $12.0^{+1.8}_{-1.7}$ &
$0.26_{-0.07}^{+0.14}$ & $1.1\pm 0.1$ & $0.75/31$ &  
0.33 \\
 CXO, PL (halo) & $10.0^{+2.2}_{-2.0}$ &
$0.6_{-0.3}^{+1.1}$ & $3.7\pm0.6$ & $0.86/17$ &  
0.31 \\
CXO, PL (point) & $15$ (fixed) &
$0.83_{-0.16}^{+0.19}$ & $2.7\pm0.2$ & $0.86/32$ &  
1.7 \\
CXO, PL (halo) & $15$ (fixed) &
$4.8_{-1.5}^{+2.2}$ & $5.0\pm0.3$ & $1.05/18$ &  
0.67 \\
CXO, PL (pre-outburst, halo) & $4.0^{+3.1}_{-1.9}$ &   
$5.5^{+4.0}_{-2.5}\times10^{-4}$ & $1.0_{-0.5}^{+0.8}$ & $0.72/7$ &  
$\sim$0.01 \\
CXO, PL (pre-outburst, halo) & $15$ (fixed) &
$4.7^{+3.2}_{-1.0}\times10^{-2}$ & $3.5\pm0.1$ & $0.80/8$ &
$\sim$0.03 \\
CXO (point) and RXTE, PL  & $20.7^{+2.2}_{-2.0}$ &
$4.8_{-2.0}^{+5.2}$ & $3.6_{-0.3}^{+0.4}$ & $0.77/41$ &  
2.8 \\
CXO (point) and RXTE, BB & $10.0^{+1.7}_{-1.5}$ &
$0.18_{-0.06}^{+0.10}$ & $1.15_{-0.08}^{+0.09}$ & $0.88/41$ &  
0.18 \\
  \hline
\end{tabular}
\end{center}
\tablecomments{
The uncertainties are given at 68\%
 confidence level for a single interesting parameter.  
 }
\tablenotetext{a}{Spectral flux in units of $10^{-2}$
 photons cm$^{-2}$ s$^{-1}$ keV$^{-1}$ at 1 keV.}
\tablenotetext{b}{
BB radius, in units of km$^2$.}
\tablenotetext{c}{Photon index or BB temperature in keV.}
\tablenotetext{c}{Reduced $\chi^2$ and the number of degrees of freedom..}
\tablenotetext{e}{Unabsorbed PL luminosity in the $2-10$ keV band or bolometric BB luminosity ($\pi R^2 \sigma T^4$), in units of $10^{34}$ ergs s$^{-1}$.}
\end{table}

 \begin{table}[]
\caption[]{Power-law fits to the phase resolved ACIS spectra of \sgrnos.\label{tab:phaseresspec}} \vspace{-0.5cm}
\begin{center}
\begin{tabular}{cccccccc}
\hline 
Phases & $N_{\rm H,22}$   & $\mathcal{N}$\tablenotemark{a}   & $\Gamma$\tablenotemark{b}   &  $\chi_{\nu}^2$/d.o.f.\tablenotemark{c}     \\
\hline
peak & 20.3 &
$1.5\pm0.2$ & $3.8\pm0.3$ & $0.94/40$  \\
off-peak & 20.3 &
$0.30\pm0.04$ & $3.5\pm0.2$ & $0.94/40$  \\
  \hline
\end{tabular}
\end{center}
\tablecomments{
$N_{\rm H}$ was held fixed during the fit. The uncertainties are given at 68\%
 confidence level for a single interesting parameter.  
 }
 \tablenotetext{a}{Spectral flux in units of $10^{-2}$
 photons cm$^{-2}$ s$^{-1}$ keV$^{-1}$ at 1 keV.}
 \tablenotetext{b}{Photon index.}
  \tablenotetext{c}{Reduced $\chi^2$ and the number of degrees of freedom.}
\end{table}

\clearpage

\begin{thebibliography}

\bibitem[Abdo et al.(2011)]{abdo2011} Abdo, A.~A., Ackermann,  M., Ajello, M., et al.\ 2011, arXiv:1108.1435	
\bibitem[Aharonian et al.(2005)]{aharonian2005} Aharonian, F., Akhperjanian, A.~G., Aye, K.-M., et al.\ 2005, Science, 307, 1938 
\bibitem[Barthelmy et al.(2011)]{barthelmy2011} Barthelmy, S.~D., et al.\ 2011, GRB Coordinates Network, 12259, 1
\bibitem[Bochow et al.(2011)]{bochow2011} Bochow, A., Carrigan, S., Gast H., et al.\ 2011, 32nd International Cosmic Ray Conference, Beijing 2011
\bibitem[Buccheri et al.(1983)]{buccheri1983} Buccheri, R., et al.\ 1983, A\&A, 128, 24
\bibitem[Chang et al.(2011)]{chang2011} Chang, C., Pavlov, G.~G., 
Kargaltsev, O., \& Shibanov, Y.~A.\ 2011, arXiv:1107.1819 
\bibitem[D'Elia et al.(2011)]{delia2011} D'Elia, V., et al.\ 2011, GRB Coordinates Network, 12253, 1 
\bibitem[Duncan \& Thompson(1992)]{duncanthompson1992} Duncan, R.~C., \& Thompson, C.\ 1992, \apjl, 392, L9 
\bibitem[Durant \& van Kerkwijk(2006)]{durant2006} Durant, M., \& van Kerkwijk, M.~H.\ 2006, \apj, 650, 1070 
\bibitem[Esposito et al.(2010)]{esposito2010} Esposito, P., Israel, 
G.~L., Turolla, R., et al.\ 2010, \mnras, 405, 1787 
\bibitem[Esposito et al.(2011)]{esposito2011} Esposito, P., Israel, G. L.,  Turolla, R., et al. 2011, MNRAS, 416, 205
\bibitem[G{\"o}{\u g}{\"u}{\c s} et al.(2010)]{gogus2010} 
G{\"o}{\u g}{\"u}{\c s}, E., Cusumano, G., Levan, A.~J., et al.\ 2010, 
\apj, 718, 331 
\bibitem[G{\"o}{\u g}{\"u}{\c s} \& Kouveliotou(2011a)]{gogus2011a} G{\"o}{\u g}{\"u}{\c s}, E., \& Kouveliotou, C.\ 2011a, GRB Coordinates Network, 12267, 1 
\bibitem[G{\"o}{\u g}{\"u}{\c s} et al.(2011b)]{gogus2011b} G{\"o}{\u g}{\"u}{\c s}, E., Kouveliotou, C., Kargaltsev, O., \& Pavlov, G.\ 2011b, GRB Coordinates Network, 12302, 1
\bibitem[Guiriec et al.(2011)]{guiriec2011} Guiriec, S., Kouveliotou, C., \& van der Horst, A.~J.\ 2011, GRB Coordinates Network, 12255, 1
\bibitem[Jahoda et al.(1996)]{jahoda1996} Jahoda, K., Swank,  J.~H., Giles, A.~B., et al.\ 1996, \procspie, 2808, 59 
\bibitem[Halpern(2011)]{halpern2011} Halpern, J.\ 2011, GRB Coordinates Network, 12260, 1
\bibitem[Hoversten et al.(2011)]{hoversten2011} Hoversten, E.~A., et al.\ 2011, GRB Coordinates Network, 12316, 1
\bibitem[Hurley(2000)]{hurley2000} Hurley, K.\ 2000, in Kippen, R. M. and Mallozzi, R. S. and Fishman, G. J.,eds., Gamma-ray Bursts, 5th Huntsville Symposium. Vol. 526 of AIP Conf.Proc., Melville NY, p. 763
\bibitem[Kargaltsev et al.(2011)]{kargaltsev2011} Kargaltsev, O., G{\"o}{\u g}{\"u}{\c s}, E., Kouveliotou, C., \& Pavlov, G.\ 2011, The Astronomer's Telegram, 3600, 1 
\bibitem[Kargaltsev \& Pavlov (2008)]{kargaltsev2008} Kargaltsev, O., \& Pavlov, G.\ G. 2008, in 40 YEARS OF PULSARS: Millisecond Pulsars, Magnetars and More. AIP Conf. Proc. 983, 171
\bibitem[Kuiper \& Hermsen(2011)]{kuiper2011} Kuiper, L., \& Hermsen, W.\ 2011, The Astronomer's Telegram, 3577, 1
\bibitem[Leahy \& Tian(2008)]{leahy2008} Leahy, D.~A., \& Tian, W.~W.\ 2008, \aj, 135, 167 
\bibitem[Levan \& Tanvir(2011)]{levan2011} Levan, A.~J., \& Tanvir, N.~R.\ 2011, GRB Coordinates Network, 12266, 1 
 \bibitem[Lucas et al.(2008)]{lucas2008} Lucas, P.~W., et al.\ 2008, \mnras, 391, 136
\bibitem[Misanovic et al.(2011)]{misanovic2011} Misanovic, Z., Kargaltsev, O., \& Pavlov, G.~G.\ 2011, \apj, 735, 33 
\bibitem[Moskvitin et al.(2011)]{moskvitin2011} Moskvitin, A.~S., Sokolov, V.~V., \& Uklein, R.~I.\ 2011, GRB Coordinates Network, 12254, 1 
\bibitem[Mukherjee et al.(2009)]{mukherjee2009} Mukherjee, R., Gotthelf, E.~V., \& Halpern, J.~P.\ 2009, \apj, 691, 1707
\bibitem[Ng et al.(2011)]{ng2011} Ng, C.-Y., Kaspi, V.~M., Dib, R., et al.\ 2011, \apj, 729, 131 
\bibitem[Nobili et al.(2008)]{nobili2008} Nobili, L., Turolla, R., \& Zane, S.\ 2008, \mnras, 389, 989 
\bibitem[Paczy{\'n}ski(1992)]{paczynski1992} Paczy{\'n}ski, B.\ 1992, Acta Astronomica, 42, 145
 \bibitem[Pavlov et 
al.(1994)]{pavlov1994} Pavlov, G.~G., Shibanov, Y.~A., Ventura, J., \& Zavlin, V.~E.\ 1994, \aap, 289, 837 
\bibitem[Pavlov et al.(1995)]{pavlov1995} Pavlov, G.~G., Shibanov, 
Y.~A., Zavlin, V.~E., 
\& Meyer, R.~D.\ 1995, The Lives of the Neutron Stars, 71 
\bibitem[Shaver 
\& Goss(1970)]{shaver1970} Shaver, P.~A., \& Goss, W.~M.\ 1970, Australian Journal of Physics Astrophysical Supplement, 14, 133  
\bibitem[Tello et al.(2011)]{tello2011} Tello, J.~C., Sota, A., \& Castro-Tirado, A.~J.\ 2011, GRB Coordinates Network, 12272, 1 
\bibitem[Tian et al.(2007)]{tian2007} Tian, W.~W., Li, Z., Leahy, D.~A., \& Wang, Q.~D.\ 2007, \apjl, 657, L25 
\bibitem[Tiengo et al.(2002)]{tiengo2002} Tiengo, A., G\"{o}hler, E., Staubert, R., \& Mereghetti, S. 2002, A\&A, 383, 182
\bibitem[Thompson \& Duncan(1995)]{thompsonduncan1995} Thompson, C., \& Duncan, R.~C.\ 1995, \mnras, 275, 255 
\bibitem[Thompson \& Duncan(1996)]{thompsonduncan1996} Thompson, C., \& Duncan, R.~C.\ 1996, \apj, 473, 322 
\bibitem[Wilson et al.(2003)]{wilson2003} Wilson, J. C., et al. 2003, Proc. SPIE, 4841, 451
\bibitem[Woods \& Thompson(2006)]{woods2006} Woods, P.~M., \& Thompson, C.\ 2006, Compact stellar X-ray sources, 547 
\bibitem[Zavlin et al.(1995)]{zavlin1995} Zavlin, V.~E., Shibanov, 
Y.~A., \& Pavlov, G.~G.\ 1995, Astronomy Letters, 21, 149 


\end{thebibliography}
\end{document}